\documentclass[aps,twocolumn]{revtex4-1} 
\usepackage[toc,page]{appendix}
\usepackage[british]{babel}
\usepackage[utf8]{inputenc}
\usepackage{amsmath,amssymb}
\usepackage{graphicx}
\usepackage{ae}
\usepackage{esdiff}
\usepackage{braket}
\usepackage{bbm}
\usepackage{simplewick}
\usepackage{float}
\usepackage{braket}
\usepackage{simplewick}
\usepackage{color}
\usepackage{framed}
\usepackage[pdftex,colorlinks=true,linkcolor=blue,citecolor=blue,urlcolor=black]{hyperref}
\usepackage{mathtools}
\usepackage{enumitem}
\usepackage{ragged2e}
\usepackage{comment}
\usepackage{import}
\usepackage[caption=false]{subfig}

\def \k{{\textbf{k}}}

\def \Q{{\textbf{Q}}}

\def \0{{\textbf{{0}}}}

\begin{document}

\title{Fermi-edge exciton-polaritons in doped semiconductor microcavities with finite hole mass}
\author{Dimitri Pimenov}
\email{D.Pimenov@physik.lmu.de}
\author{Jan von Delft}
\affiliation{Arnold Sommerfeld Center for Theoretical Physics, Ludwig-Maximilians-University Munich, 80333 Munich, Germany}
\author{Leonid Glazman}
\affiliation{Departments of Physics, Yale University, New Haven, Connecticut 06520, USA}
\author{Moshe Goldstein}
\affiliation{Raymond and Beverly Sackler School of Physics and Astronomy, Tel Aviv University, Tel Aviv 6997801, Israel}

\begin{abstract}
The coupling between a 2D semiconductor quantum well and an optical cavity gives rise to combined light-matter excitations, the exciton-polaritons. These were usually measured when the conduction band is empty, making the single polariton physics a simple single-body problem. The situation is dramatically different in the presence of a finite conduction band population, where the creation or annihilation of a single exciton involves a many-body shakeup of the Fermi sea. Recent experiments in this regime revealed a strong modification of the exciton-polariton spectrum.
Previous theoretical studies concerned with nonzero Fermi energy
mostly relied on the approximation of an immobile valence band hole with infinite mass,
which is appropriate for low-mobility samples only; for high-mobility
samples, one needs to consider a mobile hole with large but finite mass.
To bridge this gap we present
an analytical diagrammatic approach and tackle a
model with short-ranged (screened) electron-hole interaction, studying
it in two complementary regimes. We find that the finite hole mass has opposite effects on the exciton-polariton spectra in the two regimes: In the first, where the Fermi energy is much smaller than the exciton binding energy, excitonic features are  enhanced by the finite mass. In the second regime, where the Fermi energy is much larger than the exciton binding energy, finite mass effects cut off the excitonic features in the
polariton spectra, in qualitative agreement with recent experiments.
\end{abstract}
\maketitle


\section{Introduction}

When a high-quality direct semiconductor 2D quantum well (QW) is placed inside an optical microcavity, the strong coupling of photons and QW excitations gives rise to a new quasiparticle: the polariton.
The properties of this fascinating half-light, half-matter particle strongly depend on the nature of the involved matter excitations.

If the Fermi energy is in the semiconductor band gap, the matter excitations are excitons. This case is theoretically well understood  \cite{carusotto2013quantum,byrnes2014exciton}, and the first observation of the
resulting microcavity exciton-polaritons was already accomplished in 1992 by Weisbuch
\textit{et al.}\ \cite{PhysRevLett.69.3314}. Several studies on exciton-polaritons revealed remarkable results. For example, exciton-polaritons can form a Bose-Einstein condensate \cite{Kasprzak2006}, and were proposed as a
platform for high-$T_c$ superconductivity \cite{PhysRevLett.104.106402}.

The problem gets more involved if the Fermi energy is above the conduction band bottom, i.e., a conduction band Fermi sea is present. Then the matter excitations have a complex many-body structure, arising from the complementary phenomena of Anderson orthogonality~\cite{Anderson1967} and the Mahan exciton effect, entailing the Fermi-edge singularity~\cite{PhysRev.163.612,PhysRev.178.1072,NOZIERES1969,
PhysRev.178.1097,combescot1971infrared}.
An experimental study of the resulting ``Fermi-edge polaritons'' in a GaAs QW was first conducted in 2007
by Gabbay \textit{et al.}~\cite{PhysRevLett.99.157402}, and subsequently extended by Smolka \textit{et al.}~\cite{Smolka} (2014). A similar experiment on transition metal dichalcogenide monolayers was recently published by Sidler \textit{et al.}~\cite{sidler2017fermi} (2016).

From the theory side, Fermi-edge polaritons have been investigated in Ref.~\cite{PhysRevB.76.045320, PhysRevB.89.245301}. However, in these works only the case of infinite valence band hole mass was considered, which is the standard assumption in the Fermi-edge singularity or X-ray edge problem.  Such a model is valid for low-mobility samples only and thus  fails to explain the experimental findings in~\cite{Smolka}: there, a high-mobility sample was
studied, for which an almost complete vanishing of the polariton splitting was reported. Some consequences of a finite hole mass for polaritons were considered in a recent treatment~\cite{baeten2015mahan}, but without fully accounting for the so-called crossed diagrams that describe the Fermi sea shakeup, as we further elaborate below.

The aim of the present paper is therefore to study the effects of both finite mass and Fermi-edge singularity on polariton spectra in a systematic fashion. This is done analytically for a simplified model involving a contact interaction, which nethertheless preserves the qualitative features of spectra stemming from the finite hole mass and the presence of a Fermi sea. In doing so, we distinguish two regimes, with the Fermi energy $\mu$ being either much smaller or much larger than the exciton binding energy $E_B$. 
For the regime where the Fermi energy is much larger than the exciton binding energy, $\mu \gg E_B$, several treatments of finite-mass effects on the Fermi-edge singularity alone (i.e., without polaritons) are available, both analytical and numerical. Without claiming completeness, we list~\cite{gavoret1969optical,PhysRevB.44.3821,
PhysRevLett.65.1048,PhysRevB.35.7551, Nozi`eres1994}. In our work we have mainly followed the approach of
Ref.~\cite{gavoret1969optical}, extending it by going from 3D to 2D and, more importantly, by addressing the cavity coupling which gives rise to polaritons.
For infinite hole mass the sharp electronic spectral feature caused by the Fermi edge singularity can couple with the cavity mode to create sharp polariton-type spectral peaks~\cite{PhysRevB.76.045320, PhysRevB.89.245301}.
We find that the finite hole mass cuts off the Fermi edge singularity and suppresses these polariton features.

In the opposite regime of $\mu \ll E_B$, where the Fermi energy is much smaller than the exciton binding energy, we are not aware of any previous work addressing the modification of the Fermi-edge singularity due to finite mass. 
Here, we propose a way to close this gap using a diagrammatic approach. Interestingly, we find that in this regime the excitonic singularities are not cut off, but are rather enhanced by finite hole mass, in analogy to the heavy valence band hole propagator treated in~\cite{PhysRevLett.75.1988}. 

This paper has the following structure: First, before embarking into technical details, we will give an intuitive overview of the main results in Sec.~\ref{Pisummarysec}. Detailed computations will be performed in subsequent sections: 
In Sec.~\ref{Model sec}, the full model describing the coupled cavity-QW system is presented. The key quantity that determines its optical properties is the cavity-photon self-energy $\Pi$, which we will approximate by the electron-hole correlator in the absence of a cavity. Sec.~\ref{Photon self-energy zero mu sec} shortly recapitulates how $\Pi$ can be obtained in the regime of vanishing Fermi energy, for infinite and finite hole masses. Then we turn to the many-body problem in the presence of a Fermi sea in the regimes of small (Sec.~\ref{Photon self-energy small mu sec}) and large Fermi energy (Sec.\ref{Photon self-energy large mu sec}). Using the results of the previous sections, polariton properties are addressed in Sec.~\ref{Polariton properites sec}. Finally, we summarize our findings and list several possible venues for future study in Sec.~\ref{Conclusion sec}.


\section{Summary of results} 
\label{Pisummarysec}

In a simplified picture, polaritons arise from the hybridization of two quantum excitations with energies close to each other, the cavity photon and a QW resonance~\cite{carusotto2013quantum,byrnes2014exciton}. The resulting energy spectrum consists of two polariton branches with an avoided crossing, whose light and matter content are determined by the energy detuning of the cavity mode from the QW mode. 

While the cavity photon can be approximated reasonably by a bare mode with quadratic dispersion and a Lorentzian broadening due to cavity losses, the QW resonance has a complicated structure of many-body origin. The QW optical response function is rather sensitive to nonzero density of conduction band (CB) electrons.  Roughly, it tends to broaden QW spectral features, which contribute to the spectral width of polariton lines.  

A more detailed description of the polariton lines requires finding first the optical response function $\Pi(\Q, \Omega)$ of the QW alone (without polaritons).
Here, $\Q$ and $\Omega$ are, respectively, the momentum and the energy of an incident photon probing the optical response. The imaginary part of $\Pi(\Q,\Omega)$, $A(\Q, \Omega) = -\text{Im}\left[\Pi(\Q, \Omega)\right]/\pi$, defines the spectral function of particle-hole excitations in the QW. In the following, we discuss the evolution of $A(\Q, \Omega)$ as the chemical potential $\mu$ is varied, concentrating on the realistic case of a finite ratio of the electron and hole masses. We assume that the temperature is low, and consider the zero-temperature limit in the entire work. In addition, we will limit ourselves to the case where the photon is incident perpendicular to the QW, i.e.\ its in-plane momentum is zero, and study $A(\Omega) \equiv A(Q=0, \Omega)$.  

\begin{figure}[H]
\centering
\includegraphics[width=\columnwidth]{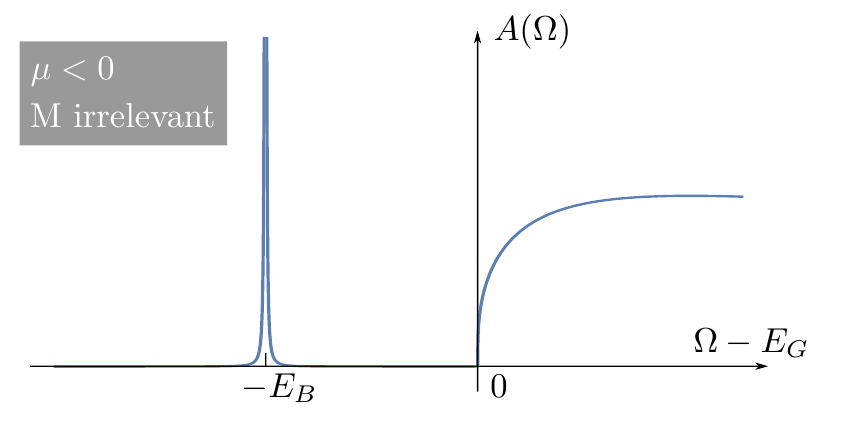}
\caption{(Color online) Absorption spectrum for short-range electron-hole interaction and $\mu<0$, given by the imaginary part of Eq.\ (\ref{ladderseries}). }
\label{mahanexciton1}
\end{figure}

\begin{figure}[H]
\centering
\includegraphics[width=\columnwidth]{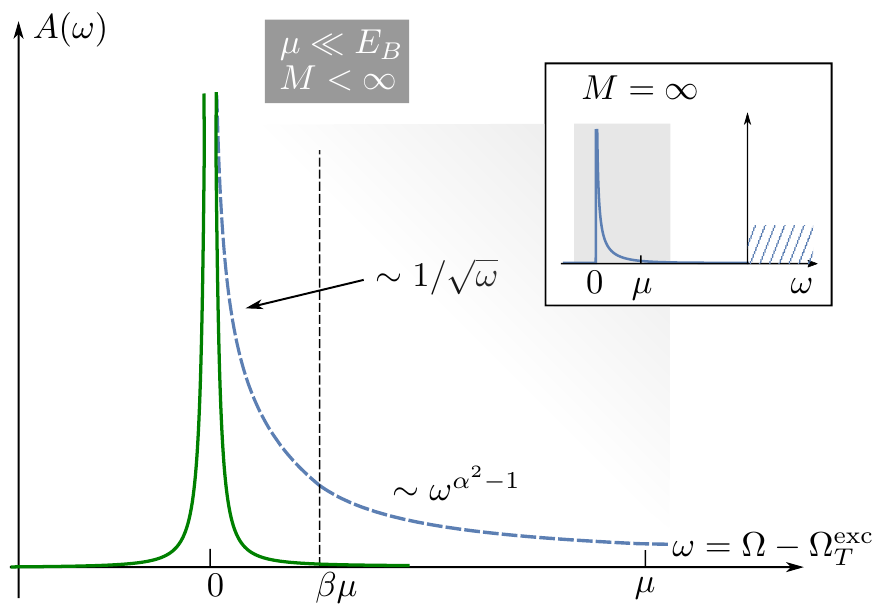}
\caption{(Color online) Absorption for $\mu \ll E_B$ and finite hole mass, illustrating Eq.\ (\ref{bothfives}). The full green curve shows the delta peak (broadened for clarity), while the dashed blue line is the incoherent part. Frequencies are measured from the exciton threshold frequency $\Omega_T^{\text{exc}} = E_G + \mu - E_B$. The inset shows the infinite mass spectrum for comparison. The dashed region in the inset indicates the continuous part of the spectrum, whose detailed form is beyond the scope of this paper, as we only consider the leading singular parts of all spectra.}
\label{finmasssmallmu1}
\end{figure}

In the absence of free carriers ($\mu$ is in the gap), a CB electron and a hole in the valence band (VB) create a hydrogen-like spectrum of bound states. In the case of a QW it is given by the 2D Elliot formula (see, e.g.,~\cite{haug1990quantum}). Being interested in the spectral function close to the main exciton resonance, we replace the true Coulomb interaction by a model of short-ranged interaction potential of strength $g$ [see Eqs.~(\ref{Helectronic}) and (\ref{gdef})].
As a result, there is a single bound state at an energy $E_G - E_B(g)$, which we identify with the the lowest-energy exciton state. Here, $E_G$ is the VB-CB gap, and energies are measured with respect to the minimum of the conduction band.
A sketch of $A(\Omega)$ is shown in Fig.~\ref{mahanexciton1}.
For $\mu>0$, electrons start to populate the CB. If the chemical potential lies within the interval $0<\mu \ll E_B$, then the excitonic Bohr radius $r_B$ remains small compared to the Fermi wavelength $\lambda_F$ of the electron gas, and the exciton is well defined. Its interaction with the particle-hole excitations in the CB modifies the spectral function $A(\Omega)$ in the vicinity of the exciton resonance. The limit of an infinite hole mass was considered 
%
by Nozi\`{e}res \textit{et al.}~\cite{PhysRev.178.1072, NOZIERES1969, PhysRev.178.1097}: Due to particle-hole excitations of the CB Fermi sea, which can happen at infinitesimal energy cost, the exciton resonance is replaced by a power law spectrum, see inset of Fig.\ \ref{finmasssmallmu1}. 
In terms of the detuning from the exciton threshold,
\begin{align}
  \omega = \Omega - \Omega_T^{\text{exc}} \ , \quad  \Omega_T^{\text{exc}} = E_G + \mu - E_B,
\end{align}
the spectral function, $A_{\text{exc}}(\omega) = - \text{Im}\left[\Pi_{\text{exc}}(\omega)\right]/\pi$, scales as: 
\begin{align}
\label{Aexcsummary}
A_{\text{exc}}(\omega)\bigg|_{M = \infty} \sim \theta(\omega) \frac{E_B}{\omega} \left(\frac{\omega} {\mu}\right)^{\alpha^2 }, \quad
\omega \ll \mu.
\end{align}
The effective exciton-electron interaction parameter $\alpha$ was found by Combescot \textit{et al.}~\cite{combescot1971infrared}, making use of final-state Slater determinants.
In their work, $\alpha$ is obtained in terms of the scattering phase shift $\delta$ of Fermi level electrons off the hole potential, in the presence of a bound state, as $\alpha = |\delta/\pi-1|$. For the system discussed here this gives \cite{adhikari1986quantum}:
\begin{align}
\alpha = 1/\left|\ln\left(\frac{\mu}{E_B}\right)\right|.
\label{excitonleadingbehaviour}
\end{align}

We re-derive the result for $\alpha$ diagrammatically (see Sec.~\ref{Photon self-energy small mu sec}), in order to extend the result of Combescot \textit{et al.}\ to the case of a small but nonzero CB electron-VB hole mass ratio $\beta$, where
\begin{align}
\beta = m/M.
\end{align}
While the deviation of $\beta$ from zero does not affect the effective interaction constant $\alpha$, it brings qualitatively new features to $A(\Omega)$, illustrated in Fig.\ \ref{finmasssmallmu1}. The origin of these changes is found in the kinematics of the interaction of the exciton with the CB electrons. Momentum conservation for finite exciton mass results in phase-space constraints for the CB particle-hole pairs which may be excited in the process of exciton creation. As a result, the effective density of states $\nu(\omega)$ of the pairs with pair energy $\omega$ (also corresponding to the exciton decay rate) is reduced from $\nu(\omega) \sim \omega$ at $\beta=0$ \cite{combescot1971infrared} to $\nu(\omega) \sim \omega^{3/2}$ when $\omega$ is small compared to the recoil energy $E_R=\beta\mu$. A smaller density of states for pairs leads to a reduced transfer of the spectral weight to the tail; therefore, the delta function singularity at the exciton resonance survives the interaction with CB electrons, i.e.\ $\beta >0$ tends to restore the exciton pole, and one finds: 
\begin{subequations}
\label{bothfives}
\begin{align}
\label{Excgeneral}
&A_{\text{exc}}(\omega)\bigg|_{M<\infty} = A_{\text{exc,incoh.}}(\omega) \theta(\omega) + \beta^{\alpha^2} E_B \delta(\omega),
\\ \notag \\
&A_{\text{exc,incoh.}}(\omega) \sim E_B
\label{Exccases}
\begin{cases}
 \frac{\alpha^2}{\sqrt{\omega \beta\mu}} \beta^{\alpha^2}  \quad & \omega \ll \beta\mu \\ 
\frac{\alpha^2}{\omega} \left(\frac{\omega}{\mu}\right)^{\alpha^2} \quad  &\beta\mu \ll \omega \ll \mu.
\end{cases}
\end{align} 
\end{subequations}
The main features of this spectral function are summarized in Fig.\ \ref{finmasssmallmu1}: 
As expected, the exciton recoil only plays a role for small frequencies $\omega \ll \beta\mu$, while the infinite mass edge singularity is recovered for larger frequencies. The spectral weight of the delta peak is suppressed by the interaction. For $\beta \rightarrow 0$ and $\alpha \ne 0$, we recover the infinite mass result, where no coherent part shows up. If, on the opposite, $\alpha^2 \rightarrow 0$ but $\beta \neq 0$, the weight of the delta peak goes to one: The exciton does not interact with the Fermi sea, and its spectral function becomes a pure delta peak, regardless of the exciton mass. A partial survival of the coherent peak at $\alpha,\beta\neq 0$ could be anticipated from the results of Rosch and Kopp~\cite{PhysRevLett.75.1988} who considered the motion of a heavy particle in a Fermi gas of light particles. This problem was also analyzed by Nozi\`eres \cite{Nozi`eres1994}, and the coherent peak can be recovered by Fourier transforming his time domain result for the heavy particle Green's function. 

At this point, let us note the following: for $\mu > 0$, the hole can bind two electrons with opposite spin, giving rise to trion features in the spectrum. We will not focus on those, since, for weak doping, their spectral weight is small in $\mu$ (more precisely, in $\mu/E_T$, where $E_T \ll E_B$ is the trion binding energy), and they are red detuned w.r.t.\ the spectral features highlighted in this work. In the regime of $\mu\gg E_B \gg E_T$, trions should be neglible as well. Some further discussion of trion properties can be found in Appendix \ref{trion-contribution}.

Upon increase of chemical potential $\mu$,
the CB continuum part (inset of Fig.\ \ref{finmasssmallmu1}) starts building up into the well-known Fermi-edge singularity
(FES) at the Burstein-Moss \cite{PhysRev.93.632,moss1954interpretation} shifted 
threshold, $\Omega_T^{
\text{FES}} = E_G + \mu$. For finite mass ($\beta \neq 0$), the FES will however be broadened by recoil effects (see below). At the same time, the delta function singularity of Eq.\ (\ref{Excgeneral}) at the absorption edge vanishes at some value of $\mu$. So, at higher electron densities, it is only the FES which yields a nonmonotonic behavior of the absorption coefficient, while the absorption edge is described by a converging power law with fixed exponent, see Eq.\ (\ref{AFES}). This evolution may be contrasted to the one at $\beta=0$. According to \cite{combescot1971infrared,PhysRevB.35.7551}, the counterparts of the absorption edge and broadened FES are two power law nonanalytical points of the spectrum which are present at any $\mu$ and characterized by exponents continuously evolving with $\mu$. 
A more detailed discussion of the evolution of absorption spectra as $\mu$ increases from small to intermediate to large values is presented in Appendix \ref{muincapp}.

\begin{figure}[H]
\centering
\includegraphics[width=\columnwidth]{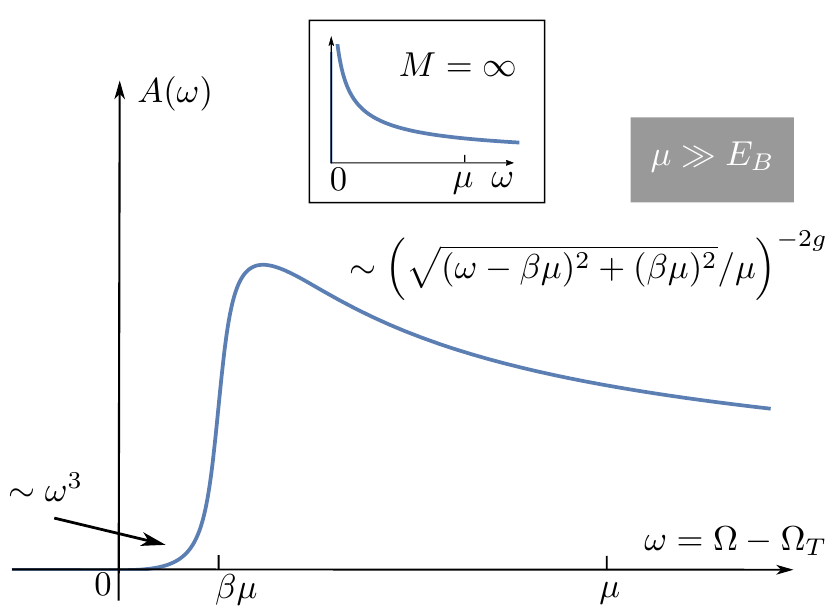}
\caption{(Color online) Finite mass absorption in the case $E_B \ll \mu$. Frequencies are measured from $\Omega_T^{\text{FES}} = E_G + \mu$. The inset shows the infinite mass case for comparison.}
\label{FEScomp}
\end{figure}

Let us now consider the limit $\mu\gg E_B$, where the FES is the most prominent spectral feature, in closer detail. In the case of infinite hole mass ($\beta=0$), and in terms of the detuning from the FES threshold, 
\begin{align}
\omega = \Omega - \Omega_T^{\text{FES}}, \quad  \Omega_T^{\text{FES}} = E_G + \mu,  
\end{align}
the FES absorption scales as \cite{PhysRev.178.1072, NOZIERES1969, PhysRev.178.1097}: 
\begin{align}
\label{FESfirst}
A_{\text{FES}}(\omega)\bigg|_{M = \infty} \sim \theta(\omega) \left(\frac{\omega}{\mu}\right)^{-2g},  \end{align}
as illustrated in the inset of Fig.\ \ref{FEScomp}.
In the above formula, the interaction contribution to the treshold shift, which is of order $g\mu$, is implicitly contained in a renormalized gap $E_G$.

What happens for finite mass? This question was answered in \cite{gavoret1969optical,PhysRevB.35.7551, Nozi`eres1994}: As before, the recoil comes into play, effectively cutting the logarithms contributing to 
(\ref{FESfirst}). Notably, the relevant quantity is now the \emph{VB hole} recoil, since the exciton is no longer a well defined entity. 
The FES is then replaced by a rounded feature, sketched in Fig.\ \ref{FEScomp}, which sets in continuously:
\begin{align}
\label{AFES}&A_\text{FES}(\omega)\bigg|_{M<\infty} \hspace{-1em} \sim
\begin{cases}
\left(\!\frac{\omega}{\beta\mu}\right)^3 \beta^{-2g} \cdot \theta(\omega) &  \omega \ll \beta\mu \\ 
\ \  \left(\frac{\sqrt{(\omega - \beta\mu)^2 + (\beta\mu)^2}}{\mu}\right)^{\!-2g}     &\beta\mu \ll \omega \ll \mu.
\end{cases}
\end{align}
Eq.\ (\ref{AFES}) can be obtained by combining and extending to 2D the results presented in Refs. \cite{gavoret1969optical,PhysRevB.35.7551}.

\begin{figure}[H]
\centering
\includegraphics[width=\columnwidth]{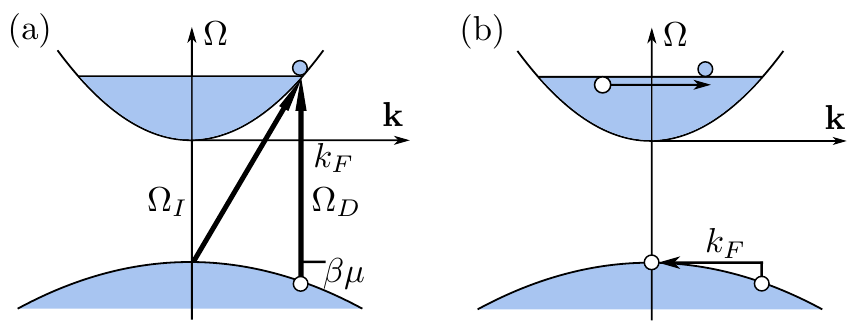}
\caption{(Color online) (a): The direct threshold $\Omega_D = \Omega_T^{\text{FES}} + \beta\mu$ and the indirect threshold $\Omega_I = \Omega_T^{\text{FES}}$ [in the main text, $\omega_{D\!/\!I} = \Omega_{D\!/\!I} - \Omega_T^{\text{FES}}]$ (b): The VB hole can undergo inelastic processes which reduces its energy, smearing the infinite mass edge singularity.}
\label{twothresholds}
\end{figure}

The maximum of Eq.\ (\ref{AFES}) is found at the so-called direct threshold, $\omega_D = \beta\mu$ (see Fig.\ \ref{twothresholds}(a)). This shift is a simple effect of the Pauli principle:  the photoexcited electron needs to be placed on top of the CB Fermi sea. The VB hole created this way, with momentum $k_F$, can subsequently decay into a zero momentum hole, scattering with conduction band electrons [see Fig.\ \ref{twothresholds}(b)]. These processes render the lifetime of the hole finite, with a decay rate $\sim g^2 \beta\mu$.
Within the logarithmic accuracy of the Fermi edge calculations, this is equal to $\beta \mu$, the cutoff of the power law in Eq.~(\ref{AFES}) (See Sec.~\ref{FESfiniteholemasssubseq} for a more detailed discussion).
As a result, the true threshold of absorption is found at the indirect threshold, $\omega_I = 0$.
Due to VB hole recoil, the CB hole-electron pair density of states now scales as $\nu(\omega) \sim\omega^3$, leading to a similar behavior of the spectrum, see Fig.\ \ref{FEScomp}.

We note that at finite ratio $\beta = m/M$, raising the chemical potential $\mu$ from $\mu \ll E_B$ to $\mu \gg E_B$ results in a qualitative change of the threshold behavior
from a singular one of Eq.\ (\ref{Exccases}), to a converging power law, see the first line of Eq.\ (\ref{AFES}). Simultaneously, a broadened FES
feature appears in the continuum, at $\omega>0$. 
The difference in the value of the exponent in the excitonic result [Eq.~(\ref{Exccases})], as compared to the FES low-energy behavior [Eq.~(\ref{AFES}) for $\omega \ll \beta\mu$], can be understood from the difference in the kinematic structure of the excitations:  In the exciton case, the relevant scattering partners are an exciton and a CB electron-hole pair. In the FES case, one has the photoexcited electron as an additional scattering partner, which leads to further kinematic constraints and eventually results in a different low-energy power law. 

In the frequency range $\beta\mu \ll \omega \lesssim \mu$, the physics is basically the same as in the infinite hole mass case ($\beta=0$). There, the behavior near the lowest threshold (which is exciton energy for $\mu \ll E_B$ and the CB continuum for $\mu \gg E_B$) is always $\sim \omega^{(1-\delta/\pi)^2-1}=\omega^{(\delta/\pi)^2-2\delta/\pi}$. But in the first case ($\mu \ll E_B$), $\delta \sim \pi - \alpha$ is close to $\pi$ (due to the presence of a bound state), so the threshold singularity is in some sense close  to the delta peak , $\sim \text{Im}[1/(\omega+i0^+)]$, that one would have for $\mu=0$, whereas in the second case ($\mu \gg E_B$), $\delta \sim g$ is close to zero, so the threshold singularity is similar to a discontinuity.

\begin{figure}[H]
\centering
\includegraphics[width=\columnwidth]{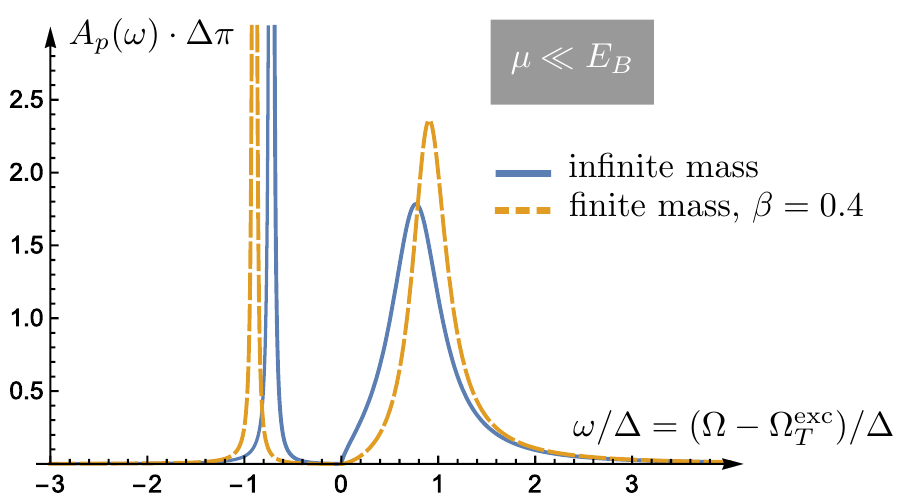}
\caption{(Color online) Comparison of the polariton spectrum for $\mu \ll E_B$, at zero cavity detuning. 
Frequencies are measured from the exciton threshold, $\Omega_T^\text{exc} = E_G + \mu-E_B$. The energy unit $\Delta$ corresponds to the half mode splitting at zero detuning in the bare exciton case ($\mu = 0$).
}
\label{summary_excpol}
\end{figure}

Having discussed spectral properties of the QW alone, we can now return to polaritons. Their spectra $A_p (\omega)$ can be obtained by inserting the QW polarization as photon self-energy. 
While a full technical account will be given in Sec.~\ref{Polariton properites sec}, the main results can be summarized as follows:

In the first case of study, of $\mu \ll E_B$ and finite $\beta$, the
polaritons arise from a mixing of the cavity and the sharp exciton mode.
The smaller the hole mass, the more singular the exciton features,
leading also to sharper polariton features. Furthermore, the enhanced
exciton quasiparticle weight pushes the two polariton branches further
apart. Conversely, in the singular limit of infinite hole mass, the pole
in the exciton spectrum turns into the pure power law familiar from
previous work, resulting in broader polariton features.  A comparison of
the infinite and finite hole mass versions of the polariton spectra
$A_p(\omega)$ when the cavity photon is tuned into resonance with the
exciton is presented in Fig.\ \ref{summary_excpol}.
Notably, the above effects are rather
weak, since the exciton is a relatively sharp resonance even for
infinite hole mass.

\begin{figure}[H]
\centering
\includegraphics[width=\columnwidth]{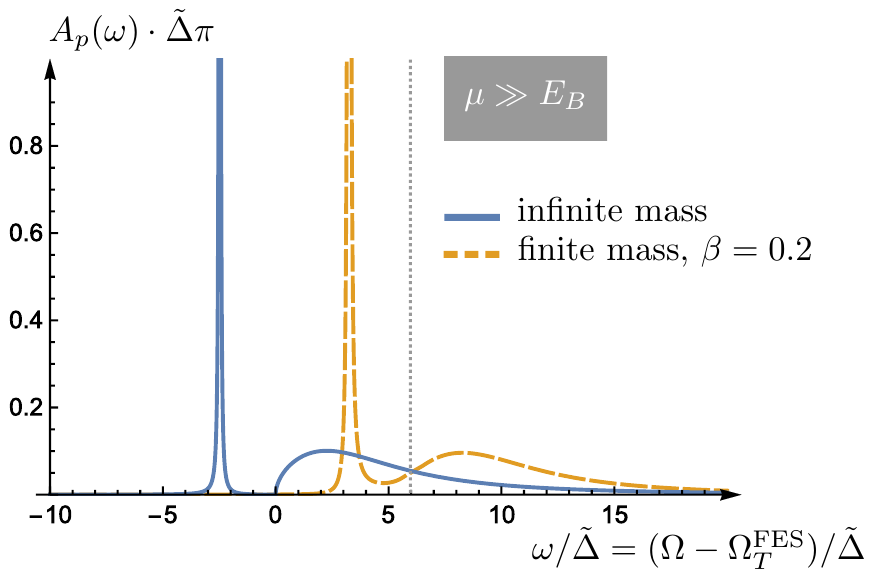}
\caption{(Color online) Comparison of the polariton spectrum for $\mu \gg E_B$, at \textbf{zero cavity detuning}. Frequencies are measured from the indirect threshold, $\Omega_T^{\text{FES}} = E_G + \mu$. The energy unit $\tilde{\Delta}$, which determines the polariton splitting at zero detuning, is defined in Sec.~\ref{Polariton properites sec}, Eq.\ (\ref{deltatildedef}). The dotted vertical line indicates the position of the direct threshold, $\omega_D = \beta\mu$.  
%
}
\label{FESpol1}
\end{figure}

In the second case, $\mu \gg E_B$, the matter component of the polaritons corresponds to the FES singularity, which is much less singular than the exciton. Consequently, the polaritons (especially the upper one, which sees the high-frequency tail of the FES) are strongly washed out already at $\beta = 0$.
For finite hole mass, the hole recoil cuts off the FES singularity, resulting in further broadening of the polaritons. In addition, there is an overall upward frequency shift by $\beta\mu$, reflecting the direct threshold effect. 
Fig.~\ref{FESpol1} shows the two polariton spectra at zero detuning.

\begin{figure}[H]
\centering
\includegraphics[width=\columnwidth]{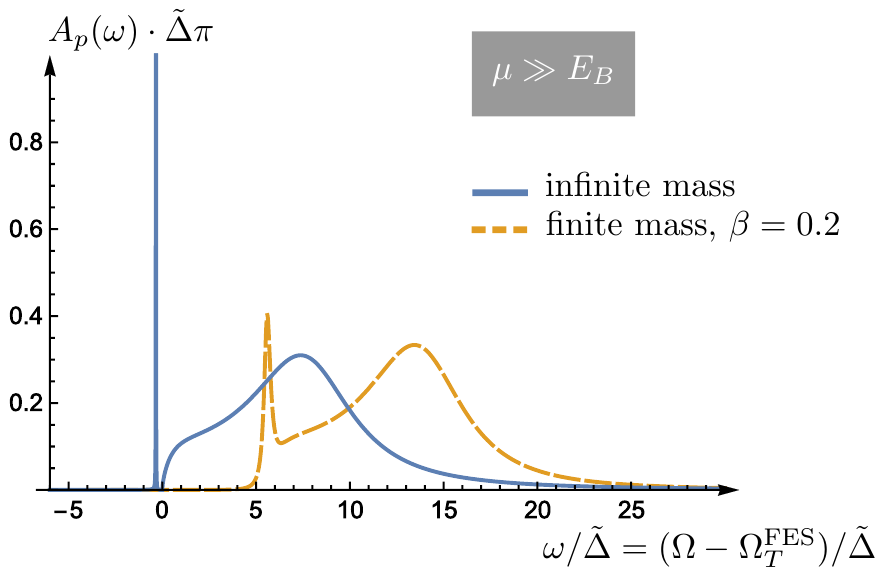}
\caption{(Color online) Comparison of the polariton spectrum for $\mu \gg E_B$, at \textbf{large positive cavity detuning}. Frequencies are measured from the indirect threshold, $\Omega_T^{\text{FES}} = E_G + \mu$. 
}
\label{FESpol2}
\end{figure}

The cutoff of the lower polariton for finite masses is even more drastic when the cavity is blue-detuned with respect to the threshold: Indeed, at large positive cavity detuning, the lower polariton is mostly matter-like, and thus more sensitive to the FES broadening. It therefore almost disappears, as seen in Fig.~\ref{FESpol2}. 


\section{Model}
\label{Model sec}

After the qualitative overview in the previous section, let us now go into more detail, starting with the precise model in question.
To describe the coupled cavity-QW system, we study the following 2D Hamiltonian: 
\begin{align}
\label{fullhamil}
H &= H_M + H_L, \\
H_M &=
\label{Helectronic} \sum_{\textbf{k}} \epsilon_\textbf{k} a^{\dagger}_{\textbf{k}}a_{\textbf{k}} - \sum_{\textbf{k}} \left[E_\textbf{k} +E_G\right] b^{\dagger}_{\textbf{k}}b_{\textbf{k}} \\ \nonumber &\quad  - \frac{V_0}{\mathcal{S}}\sum_{\textbf{k}, \textbf{p}, \textbf{q}}  a^{\dagger}_\textbf{k} a_\textbf{p} b_{\textbf{k}-\textbf{q}}b^\dagger_{\textbf{p} - \textbf{q}}, \\
H_L &=  \sum_{\textbf{Q}}\omega_{\textbf{Q}} c^\dagger_\textbf{Q} c_{\textbf{Q}}   -i\frac{d_0}{\sqrt{\mathcal{S}}}\sum_{ \textbf{p},\textbf{Q}} a^\dagger_{\textbf{p}+ \textbf{Q}}b_{\textbf{p}}c_{\textbf{Q}}
+ \text{h.c.} 
\end{align}
Here, $H_M$, adapted from the standard literature on the X-ray edge problem \cite{gavoret1969optical}, represents the matter part of the system, given by a semiconductor in a two-band approximation: $a_\textbf{k}$ annihilates a conduction band (CB) electron with dispersion $\epsilon_{\textbf{k}} = \frac{k^2}{2m}$, while $b_{\textbf{k}}$ annihilates a valence band (VB) electron with dispersion $-(E_{\textbf{k}} + E_G) = -(\frac{k^2}{2M} + E_G)$. $E_G$ is the gap energy, which is the largest energy scale under consideration: In GaAs, $E_G \simeq 2$eV, while all other electronic energies are on the order of meV. The energies are measured from the bottom of the conduction band. $\mathcal{S}$ is the area of the QW, and we work in units where $\hbar = 1$. Unless explicitly stated otherwise, we assume spinless electrons, and concentrate on the zero temperature limit. 

When a valence band hole is created via cavity photon absorption, it interacts with the conduction band electrons with an attractive Coulomb interaction. Taking into account screening, we model the interaction as point-like, with a constant positive matrix element $V_0$. The effective potential strength is then given by the dimensionless quantity
\begin{align}
\label{gdef}
g = \rho V_0,  \quad \rho = \frac{m}{2 \pi},
\end{align}
$\rho$ being the 2D DOS. The appropriate value of $g$ will be further discussed in the subsequent sections.

Interactions of CB electrons with each other are completely disregarded in Eq.~(\ref{fullhamil}), presuming a Fermi liquid picture. This is certainly a crude approximation. It can be justified if one is mostly interested in the form of singularities in the spectral function. These are dominated by various power laws, which arise from low-energy particle hole excitations of electrons close to the Fermi energy, where a Fermi-liquid description should be valid.  

The photons are described by $H_L$: We study lossless modes with QW in-plane momenta $\textbf{Q}$ and energies $\omega_\textbf{Q} = \omega_c + Q^2/2m_c$, where $m_c$ is the cavity mode effective mass. Different in-plane momenta $\textbf{Q}$ can be achieved by tilting the light source w.r.t.\ the QW. In the final evaluations we will mostly set $\textbf{Q}=0$, which is a valid approximation since $m_c$ is tiny compared to electronic masses. 
The interaction term of $H_L$ describes the process of absorbing a photon while creating an VB-CB electron hole pair, and vice versa. $d_0$ is the interband electric dipole matrix element, whose weak momentum dependence is disregarded. This interaction term can be straightforwardly derived from a minimal coupling Hamiltonian studying interband processes only, and employing the rotating wave and electric dipole approximations (see, e.g., \cite{yamamoto1999mesoscopic}). 

The optical properties of the full system are determined by the retarded dressed photon Green's function~\cite{PhysRevB.89.245301, baeten2015mahan}:
\begin{align}
\label{dressedphot}
D^R(\textbf{Q},\Omega) = \frac{1}{\Omega - \omega_\textbf{Q} + i0^+ - \Pi(\textbf{Q},\Omega)},
\end{align}
where $\Pi(\textbf{Q},\Omega)$ is the retarded photon self-energy. This dressed photon is nothing but the polariton. The spectral function corresponding to (\ref{dressedphot}) is given by 
\begin{align}
\label{Polaritonspectralfunction}
\mathcal{A}(\textbf{Q},\omega) = -\frac{1}{\pi}\text{Im}\left[D^R(\textbf{Q},\omega)\right]. 
\end{align}
$\mathcal{A}(\textbf{Q},\omega)$ determines the absorption respectively reflection of the coupled cavity-QW system, which are the quantities typically measured in polariton experiments like \cite{PhysRevLett.99.157402,Smolka}.
 
\begin{figure}[H]
\centering
\includegraphics[width=.7\columnwidth]{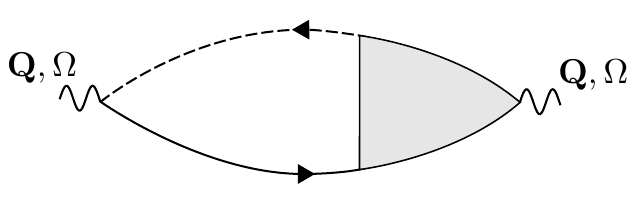}
\caption{The photon self-energy $\Pi(\textbf{Q},\Omega)$ in linear response. Full lines denote CB electrons, dashed lines VB electrons, and wavy lines photons. The grey-shaded area represents the full CB-VB vertex.}
\label{bubble}
\end{figure}

Our goal is to determine $\Pi(\textbf{Q},\Omega)$. To second order in $d_0$ it takes the form
\begin{align}
\label{Kubo-formula}
\Pi(\textbf{Q},\Omega) \simeq &-i\frac{d_0^2}{\mathcal{S}}\int_{-\infty}^{\infty}\! dt \theta(t) e^{i\Omega t}\\
\nonumber&
\times \sum_{\textbf{k},\textbf{p}}\braket{0| b_{\textbf{k}}^\dagger(t) a_{\textbf{k}+\textbf{Q}}(t) a^\dagger_{\textbf{p}+\textbf{Q}}(0) b_{\textbf{p}} (0)|0} ,
\end{align}
where $\ket{0}$ is the noninteracting electronic vacuum with a filled VB, and the time dependence of the operators is generated by $H_M$.
Within this approximation, $\Pi(\textbf{Q},\omega)$ is given by the ``dressed bubble'' shown in Fig.~\ref{bubble}.
The imaginary part of $\Pi(\textbf{Q},\omega)$ can also be seen as the linear response absorption of the QW alone with the cavity modes tuned away.

Starting from Eq.~(\ref{Kubo-formula}), in the following we will study in detail how $\Pi(\textbf{Q},\omega)$ behaves as the chemical potential $\mu$ is increased, and distinguish finite and infinite VB masses $M$. We will also discuss the validity of the approximation of calculating $\Pi$ to lowest order in $d_0$.


\section{Electron-hole correlator in the absence of a Fermi sea}
\label{Photon self-energy zero mu sec}

\begin{figure}[H]
\centering
\includegraphics[width=\columnwidth]{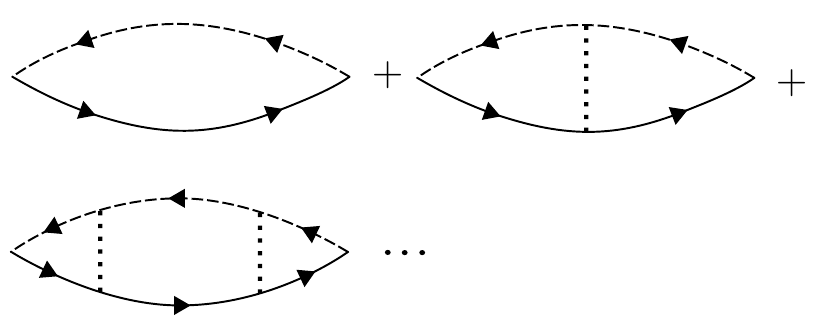}
\caption{The series of ladder diagrams. Dotted lines represent the electron-hole interaction.}
\label{excladder2}
\end{figure}

We start by shortly reviewing the diagrammatic approach in the case when the chemical potential lies within the gap (i.e. $-E_G<\mu<0$). This is mainly done in order to set the stage for the more involved diagrammatic computations in the subsequent sections.
In this regime of $\mu$, 
$\Pi$ is exactly given by the sum of the series of ladder diagrams shown in Fig.~\ref{excladder2}, first computed by Mahan \cite{PhysRev.153.882}.
Indeed, all other diagrams are absent here since they either contain VB or CB loops, which are forbidden for $\mu$ in the gap.
This is seen using the following expressions for the zero-temperature time-ordered free Green's functions: 
\begin{align}
\label{standardformula}
G_{c}^{(0)}(\textbf{k},\Omega) &= \frac{1}{\Omega - \epsilon_\textbf{k} + i0^+\text{sign}(\epsilon_\textbf{k}-\mu)},
\\ 
G_{v}^{(0)}(\textbf{k},\Omega) &= \frac{1}{\Omega + E_G + E_{\textbf{k}} + i0^+\text{sign}(-E_G-E_\textbf{k}-\mu)},
\end{align}
where the indices $c$ and $v$ stand for conduction and valence band, respectively, and $0^+$ is an infinitesimal positive constant. For $-E_G < \mu< 0$, CB electrons are purely retarded,  while VB electrons are purely advanced. Thus, no loops are possible. Higher order terms in $d_0$ are not allowed as well.

One can easily sum up the series of ladder diagrams assuming the simplified interaction $V_0$~\cite{gavoret1969optical}. Let us start from the case of infinite VB mass ($\beta=0$), and concentrate on energies $|\Omega - E_G| \ll \xi$,  where $\xi$ is an appropriate UV cutoff of order of CB bandwidth.
Since the interaction is momentum independent, all integrations in higher-order diagrams factorize. Therefore, the $n$-th order diagram of Fig. \ref{excladder2} is readily computed: 
\begin{align}
\label{ladder contribution}
\Pi^{(n)}_\text{ladder}(\Omega) = d_0^2
\rho (-g)^n\ln\left(\frac{\Omega - E_G + i0^+}{ - \xi }\right)^{n+1}.
\end{align}
Here and henceforth, the branch cut of the complex logarithm and power laws is chosen to be on the negative real axis. 
%
The geometric series of ladder diagrams can be easily summed:
\begin{align}
\label{ladderseries}
\Pi_{\text{ladder}}(\Omega) = \sum_{n=0}^{\infty} \Pi^{(n)}_{\text{Ladder}}(\Omega) = \frac{d_0^2\rho \ln\left(\frac{\Omega-E_G + i0^+}{-\xi}\right)}{1+g\ln\left(\frac{\Omega-E_G + i0^+}{-\xi}\right)}.
\end{align}
A sketch of the corresponding QW absorption $A_{\text{ladder}}= -\text{Im}[\Pi_{\text{ladder}}]/\pi$ was already shown in Fig.~\ref{mahanexciton1}.

$\Pi_{\text{ladder}}(\Omega)$ has a pole,
the so-called Mahan exciton \cite{PhysRev.153.882, gavoret1969optical}, at an energy of 
\begin{align}
\label{EBfctofg}
\Omega - E_G = -E_B = -\xi e^{-1/g}. 
\end{align}
In the following, we will treat $E_B$ as a phenomenological parameter. To match the results of the short-range interaction model with an experiment, one should equate $E_B$ with $E_0$, the energy of lowest VB hole-CB electron hydrogenic bound state (exciton).
Expanding Eq.~(\ref{ladderseries}) near the pole, we obtain: 
\begin{align}
\label{infmassmahanexciton}
\Pi_{\text{ladder}}(\omega) &= \frac{d_0^2E_B \rho}{g^2} G^0_{\text{exc}}(\omega) + \mathcal{O}\left(\frac{\omega}{E_B}\right),\\ \nonumber  G^0_{\text{exc}}(\omega) &= \frac{1}{\omega+i0^+},
\end{align}
where $\omega = \Omega-E_G + E_B$, and we have introduced the bare exciton Green's function $G^0_{\text{exc}}$, similar to Ref. \cite{Betbeder-Matibet2001}.

In this regime of $\mu$, a finite hole mass only results in a weak renormalization of the energy by factors of $1+\beta$, where $\beta= m/M$ is the small CB/VB mass ratio. Furthermore, if finite photon momenta $\textbf{Q}$ are considered, the exciton Green's function is easily shown to be (near the pole):
\begin{align}
\label{finmassmahanexciton}
G_{\text{exc}}^{0}(\textbf{Q},\omega) = \frac{1}{\omega + Q^2/M_\text{exc} + i0^+},
\end{align}
with $M_\text{exc} = M + m = M (1+\beta)$.


\section{Electron-hole correlator for small Fermi energy}
\label{Photon self-energy small mu sec}

\subsection{Infinite VB hole mass}

Let us now slightly increase the chemical potential $\mu$, and study the resulting absorption. More precisely, we consider the regime
\begin{align}
\label{scales1}
0<\mu \ll E_B \ll \xi.
\end{align}
We first give an estimate of the coupling constant $g = \rho V_0$ Accounting for screening of the VB hole 2D Coulomb potential by the CB Fermi sea in the static RPA approximation, and averaging %
over the Fermi surface~\cite{gavoret1969optical,PhysRev.153.882}
one finds:
\begin{align}
\label{gestimate}
g \sim
\begin{cases}
1-8x/\pi &  x\rightarrow 0,\\
\ln(x)/x  & x\rightarrow \infty,
\end{cases}
\end{align}
where $x= \sqrt{\mu/E_0}$ with $E_0$ being the true 2D binding energy of the lowest exciton in the absence of a CB Fermi sea.
In the regime under study we may assume $E_B \simeq E_0 \gg \mu$, and therefore $g \lesssim 1$ \footnote{Strictly speaking, this also means $E_B \lesssim \xi$, contradicting Eq.~(\ref{scales1}). However, this clearly is a non-universal property, and we will not pay any attention to it in the following}. As a result, perturbation theory in $g$ is meaningless. Instead, we will use $\mu/E_B$ as our small parameter, and re-sum all diagrams which contribute to the lowest nontrivial order in it.

We will now restrict ourselves to the study of energies close to $E_B$ 
in order to understand how a small density of CB electrons modifies the shape of the bound state resonance; we will not study in detail the VB continuum in the spectrum (cf.\ Fig.~\ref{finmasssmallmu1}).
We first compute the contribution of the ladder diagrams; 
as compared to Eqs.~(\ref{infmassmahanexciton})--(\ref{finmassmahanexciton}), the result solely differs by a shift of energies: 
\begin{align}
\label{muexcitonpole}
\omega = \Omega - \Omega_T^{\text{exc
}}, \quad  \Omega_T^{\text{exc}} = (E_G + \mu) - E_B.
\end{align}
Also, the continuum now sets in when $\Omega$ equals $\Omega_{T}^{\text{FES}} = E_G + \mu$, which is known as the Burstein-Moss shift \cite{PhysRev.93.632,moss1954interpretation}. 
However, for finite $\mu$ one clearly needs to go beyond the ladder approximation, and take into account the ``Fermi sea shakeup''. To do so, we first consider the limit of infinite $M$ ($\beta=0$). In this regime, the QW absorption in the presence of a bound state for the model under consideration was found by Combescot and Nozières \cite{combescot1971infrared}, using a different approach \footnote{In fact, their computation is in 3D, but the case of infinite hole mass is effectively 1D anyway.}.

For finite $\mu$, the physics of the  Fermi-edge singularity comes into play: Due to the presence of the CB Fermi sea, CB electron-hole excitations are possible at infinitesimal energy cost.

\begin{figure}[H]
\centering
\includegraphics[width=\columnwidth]{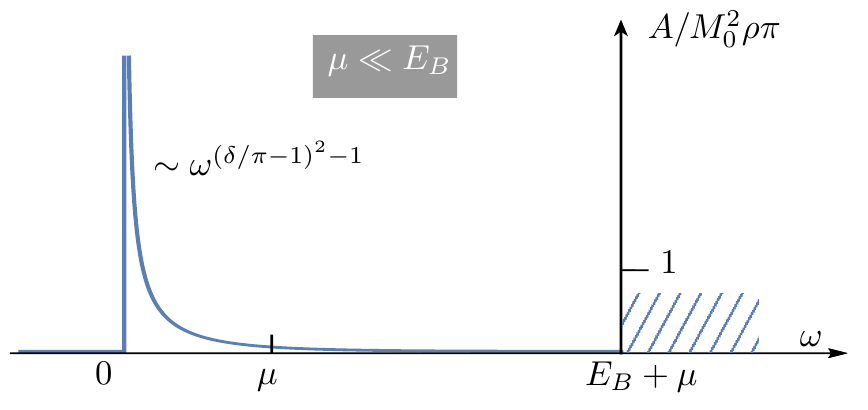}
\caption{(Color online) QW Absorption for $\mu\ll E_B$ and $M=\infty$. The power law (\ref{Nozieresresult}) is valid asymptotically close to the left peak. The dashed region indicates the continuous part of the spectrum, compare caption of Fig.\ \ref{finmasssmallmu1}.}
\label{Infmasssmallmu}
\end{figure}

As a result,
the exciton Green's function, which we analogously to (\ref{infmassmahanexciton}) define as proportional to the dressed bubble in the exciton regime,
\begin{align}
&\Pi_{\text{exc}}(\omega) = \frac{d_0^2 E_B \rho}{g^2} G_{\text{exc}}(\omega) + \mathcal{O}\left(\frac{\omega}{E_B}\right)  ,  \\   &G_{\text{exc}}(\omega) = \frac{1}{\omega - \Sigma^{\text{exc}}(\omega)}  , \label{dressedexcwithsigma}
\end{align}  
gets renormalized by a self-energy $\Sigma^{\text{exc}}(
\omega)$. This self-energy turns the 
 exciton pole turns into a divergent power law~\cite{combescot1971infrared}:
\begin{align}
\label{Nozieresresult}
G_{\text{exc}}(\omega) \sim \frac{1}{\omega+i0^+}\cdot \left(\frac{\omega+i0^+}{-\mu}\right)^{(\delta/\pi -1)^2},
\end{align}
where $\delta$ is the scattering phase shift of electrons at the Fermi-level off the point-like hole potential. One should note that no delta-peak will appear for $\delta/\pi \neq 1$. A sketch of the resulting absorption $A$ is shown in Fig.~\ref{Infmasssmallmu}.

Let us further discuss the result~(\ref{Nozieresresult}). It was obtained in~\cite{combescot1971infrared} using an elaborate analytical evaluation of final state Slater determinants, and actually holds for any value of $\mu$. A numerical version of this approach for the infinite VB mass case was recently applied by Baeten and Wouters~\cite{PhysRevB.89.245301} in their treatment of polaritons. In addition, the method was numerically adapted to finite masses by Hawrylak~\cite{PhysRevB.44.3821}, who, however, mostly considered the mass effects for $\mu\gg E_B$.

However, due to the more complicated momentum structure, it seems difficult to carry over the method of~\cite{combescot1971infrared} to finite masses analytically. Instead, we will now show how to proceed diagrammatically.
Our analysis will give (\ref{Nozieresresult}) to leading order in the small parameter $\mu/E_B$, or, equivalently, $\alpha = \delta/\pi -1$ (recall that by Levinson's theorem~\cite{adhikari1986quantum} $\delta=\pi$ for $\mu=0$ due to the presence of a bound state --- the exciton):
\begin{align}
\label{excspec}
G_{\text{exc}}(\omega) \simeq  \frac{1}{\omega+i0^+}\left(1+ \alpha^2 \ln\left(\frac{|\omega|}{\mu}\right)-i\alpha^2\pi\theta(\omega)\right).
\end{align} 
The merit of the diagrammatical computation is twofold: First, it gives an explicit relation between
$\alpha$ and the experimentally-measurable parameters $\mu$, $E_B$. Second, the approach can be straightforwardly generalized to finite masses, as we show in the next subsection.

Let us note that a similar diagrammatic method was also examined by Combescot, Betbeder-Matibet \textit{et al.}\ in a series of recent papers~\cite{Betbeder-Matibet2001,Combescot2002,combescot2003commutation,Combescot2008215,5914361120110215}. Their model Hamiltonians are built from realistic Coulomb electron-hole and electron-electron interactions. As a result, they assess the standard methods of electron-hole diagrams as too complicated \cite{Betbeder-Matibet2001}, and subsequently resort to exciton diagrams and the so-called commutation technique, where the composite nature of the excitons is treated with care. However, the interaction of excitons with a Fermi sea is only treated at a perturbative level, assuming that the interaction is small due to, e.g., spatial separation~\cite{Combescot2002}. This is not admissible in our model, where the interaction of the VB hole with all relevant electrons (photoexcited and Fermi sea) has to be treated on the same footing. Rather, we stick to the simplified form of contact interaction, and show how one can use the framework of standard electron-hole diagrams to calculate all quantities of interest for infinite as well as for finite VB mass. The results presented below then suggest that for $\mu \ll E_B$ the finite mass does not weaken, but rather strengthens the singularities, which is in line with results on the heavy hole found in~\cite{PhysRevLett.75.1988}. 
 
\begin{figure}[H]
\centering
\includegraphics[width=\columnwidth]{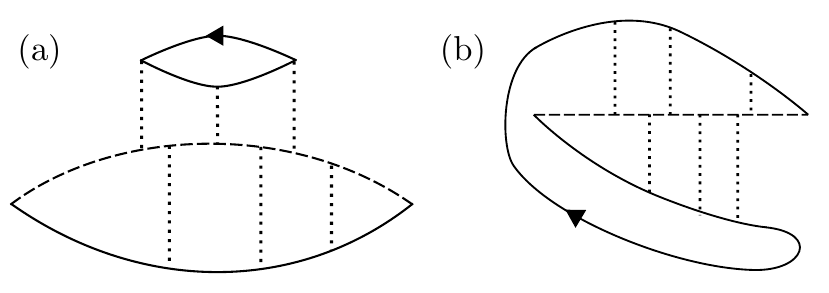}
\caption{Leading-order direct self-energy diagrams: (a) direct contribution $D$ and (b) exchange contribution $X$.}
\label{directtimedomain}
\end{figure}

Here we only present the most important physical ingredients for our approach, and defer the more technical details to Appendix~\ref{technical}.
In the regime of interest, we can perform a low-density computation, employing the small parameter $\mu/E_B$. Since all energies are close to $E_B$, the leading-order exciton self-energy diagrams is then the sum of all diagrams with one CB electron loop. One can distinguish two channels: direct and exchange, to be denoted by $D$ and $X$, as depicted in  Fig.~\ref{directtimedomain}.
All such diagrams with an arbitrary number of interactions connecting the VB line with the CB lines in arbitrary order have to be summed. Factoring out $E_B\rho/g^2 \cdot G^{0}_{\text{exc}}(\omega)^2$, the remaining factor can be identified as the exciton self-energy diagram.

An evaluation of these diagrams is possible either in the time or in the frequency domain. Of course, both approaches must give same result. In practice, however, the time domain evaluation is more instructive and requires less approximations, which is why we will discuss it first. The frequency domain evaluation, however, is far more convenient for obtaining finite mass results, and will be discussed thereafter.  

The time domain approach is similar in spirit to the classical one-body solution of the Fermi-edge problem by Nozières and de Dominicis \cite{PhysRev.178.1097}. Since the infinite-mass hole propagator is trivial, $G_v(t) = i\theta(-t)e^{i E_G t}$, the direct diagrams just describe the independent propagation of two electrons in the time-dependent hole potential. Thus, in the time domain the sum of all direct diagrams $D(t)$ factorizes into two parts representing the propagation of these two electrons:
\begin{align}
\label{Dnew}
D(t) = \int_{k_1<k_F} \frac{d\textbf{k}_1}{(2 \pi)^2} i e^{-i(E_G-\epsilon_{\textbf{k}_1})t} B(t) C(t),
\end{align}
where $B(t)$, $C(t)$ are infinite sums of convolutions (denoted by an asterisk) of the form 
\begin{align} &
B(t) = \sum_{m=1}^{\infty} (-V_0)^m \int_{k_2>k_F} \frac{d\textbf{k}_2}{(2 \pi)^2} ... \int_{k_{m}>k_F} \frac{d\textbf{k}_{m}}{(2 \pi)^2} \\ \nonumber  &\left[G_c^{0, R}(\textbf{k}_1,\ ) \ast \cdots \ast G_c^{0, R}(\textbf{k}_{m},\ )\ast G_c^{0, R} (\textbf{k}_1,\ )\right](t), 
\end{align}
and similarly for $C(t)$. $G_c^{{0},R}$ is the retarded bare CB Green's function in the time domain. Fourier-transforming, $D(\omega)$ is then given by a convolution of $B(\omega)$ and $C(\omega)$, each of which in turn reduces to simple summations of ladder diagrams. The full convolution $D(\omega)$ is difficult to compute; one can proceed by noting that $B(\omega)$, $C(\omega)$ have poles at $\omega \simeq 0$ and continuum contributions at $\omega \gtrsim E_B$. These are readily identified with the pole and continuum contributions of the exciton absorption, c.f.\ Fig.~\ref{mahanexciton1}. Combining these, there are four combinations contributing to $D(\omega)$: pole-pole, pole-continuum (two possibilities), and continuum-continuum. The imaginary part of the latter, which is of potential importance for the line shape of the exciton spectrum, can be shown to vanish in our main regime of interest, $\omega \gtrsim 0$.
It is instructive to study the pole-pole combination, which corresponds to a would be ``trion'' (bound state of the exciton and an additional electron) and is further discussed in Appendix~\ref{trion-contribution}.
Adding to it the pole-continuum contributions we find, for small $\omega$:
\begin{align}
\label{Ddirectfinal}
D(\omega) = \frac{\rho E_B}{g^2} \frac{1}{(\omega + i0^+)^2}  \Sigma_\text{exc}^{\text{D}}(\omega).
\end{align} 
This corresponds to a contribution to the exciton self-energy which reads: 
\begin{align}
\label{SigmaDint}
\Sigma^\text{D}_{\text{exc}}(\omega) = -\frac{1}{\rho}\int_{k_1<k_F} \frac{d\textbf{k}_1}{(2\pi)^2} \frac{1}{\ln\left(\frac{\omega + \epsilon_{\textbf{k}_1} - \mu + i0^+}{-E_B}\right)} . 
\end{align}
Before discussing this term further, we consider the contribution of the exchange diagrams, $X(\omega)$, of Fig.\ \ref{directtimedomain}(b). Their structure is more involved compared to the direct channel, since these diagrams do not just represent the independent propagation of two electrons in the hole potential. However, relying on a generalized convolution theorem which we prove, the computation can be performed in the same vein as before (see Appendix~\ref{technical}), leading to the following results: 
First, the pole-pole contribution cancels that of the direct diagrams (see Appendix~\ref{trion-contribution}), which holds in the spinless case only (in the spinful case, the direct diagrams will come with an extra factor of two). This could be expected: trion physics is only recovered in the spinful case, where two electrons can occupy the single bound state created by the attractive potential of the hole.
In a realistic 2D setup trion features will become important for large enough values of $\mu$ (see, e.g., \cite{sidler2017fermi,suris2001excitons, PhysRevB.91.115313,efimkin2017many}). Although we do not focus on trions here, let us stress that all standard results on trions can be recovered within our diagrammatic approach, if electrons and holes are treated as spin-$1/2$ particles; see Appendix \ref{trion-contribution} for further details.

The dominant contribution to $X(\omega)$ then arises from the pole-continuum contribution. It is given by: 
\begin{align}
\label{Xomegamaintext}
X(\omega) = -\frac{\rho E_B}{g^2} \frac{1}{(\omega + i 0^+)^2} \mu.
\end{align}
Thus, the self-energy contribution to the exciton Green's function is simply
\begin{align}
\label{Fumitypeshift}
\Sigma_{\text{exc}}^{\text{X}}(\omega) = -\mu.
\end{align} 
Since it is purely real, it will essentially just red-shift the exciton pole by $\mu$. A discussion of this result is presented in Appendix~\ref{Fumidiscussion}.

Now, it should be noted that $\Sigma_\text{exc}^{\text{X}}(\omega)$ is not proportional to the small parameter $\mu/E_B$ -- the latter effectively canceled when factoring out the bare excitons Green's function. Thus, it is inconsistent to treat $\Sigma_\text{exc}^{\text{X}}(\omega)$ as perturbative self-energy correction. Instead, one should repeat the calculation, but replace all ladders by ladders dressed with exchange-type diagrams. It can be expected, however, that the structure of the calculations will not change. The only change that should happen is the appearance of the renormalized binding energy $\tilde{E}_B = E_B + \mu$, in accordance with~\cite{combescot1971infrared}, as discussed in Appendix~\ref{Fumidiscussion}. In the following, we will assume this is accounted for, and therefore suppress all exchange diagrams.

Let us now return to the direct self-energy contribution $\Sigma_\text{exc}^{\text{D}}(\omega)$, Eq.~(\ref{SigmaDint}), writing 
\begin{align}
\Sigma_{\text{exc}}(\omega) = \Sigma^D_{\text{exc}}(\omega) 
\end{align}
henceforth. We may apply the following asymptotic expansion for the logarithmic integral (generalized from~\cite{R.Wong1989}), which will also prove useful later: 
\begin{align}
\label{theorem}
\int _0^\omega dx \frac{x^n}{\ln^m(x)} = \frac{1}{\ln^m(\omega)}\frac{\omega^{n+1}}{(n+1)} + \mathcal{O}\left(\frac{\omega^{n+1}}{\ln(\omega)^{m+1}}\right).
\end{align}
This can be shown easily by integrating by parts and comparing orders.
Based on this result we find, to leading logarithmic accuracy,
\begin{align}
\label{sigmadendlich}
\Sigma_\text{exc}(\omega) \simeq& -\frac{\mu}{\ln\left(\frac{\mu}{E_B}\right)} + \frac{\omega \ln\left(\frac{|\omega|}{\mu}\right)}{\ln\left(\frac{\mu}{E_B}\right)\!\ln\left(\frac{|\omega|}{E_B}\right)} \\&- i\frac{\pi \omega \theta(\omega)}{\ln^2\left(\frac{|\omega|}{E_B}\right)}. \notag
\end{align}
This result has several interesting features.
First, we see the appearance of a small parameter $\alpha \equiv 1/|\ln(\mu/E_B)|$, which can be interpreted as follows:
the scattering phase-shift at the Fermi level, $\delta$, which determines the Anderson orthogonality power law [c.f.\ Eq.~(\ref{Nozieresresult})] is approximately given by~\cite{adhikari1986quantum}
\begin{align}
\delta \simeq \frac{\pi}{\ln\left(\frac{\mu}{E_B}\right)} + \pi ,
\end{align}
which holds for small Fermi energies, where $\delta$ is close to $\pi$. 
Therefore, $\delta$ and $\alpha$ are related by:
\begin{align}
\label{alphaisphase}
\alpha \simeq 1-\frac{\delta}{\pi} .
\end{align}
The small pole shift of order $\alpha\mu$ contained in Eq.~(\ref{sigmadendlich}) could be expected from Fumi's theorem (see, e.g., \cite{G.D.Mah2000} and the discussion in Appendix~\ref{Fumidiscussion}). We now perform an energy shift
\begin{align}
\omega \rightarrow \omega + \alpha\mu.
\end{align}
To leading order in $\alpha$, we may then rewrite $\Sigma_\text{exc}^{\text{D}}$ with  logarithmic accuracy as
\begin{align}
\label{Sigmanice}
\Sigma_\text{exc}(\omega) \simeq \alpha^2\omega \ln\left(\frac{|\omega|}{\mu}\right) - i\alpha^2\pi \omega \theta(\omega),
\end{align}
Here, the imaginary part can be identified with the density of states of CB electron-hole excitations as function of $\omega$, as discussed in Sec.~\ref{Pisummarysec}.

Upon inserting (\ref{Sigmanice}) into the exciton Green's function (\ref{dressedexcwithsigma}), we recover (\ref{Nozieresresult}) to leading (quadratic) order in $\alpha$:
\begin{align}
\label{excspec2}
G_{\text{exc}}(\omega) \simeq  \frac{1}{\omega+i0^+}\left(1+ \alpha^2 \ln\left(\frac{|\omega|}{\mu}\right)-i\alpha^2\pi\theta(\omega)\right).
\end{align}  
As a result, our one-loop computation has given the first logarithm of the orthogonality power law, in complete analogy to the standard Fermi-edge problem (see Sec.~\ref{Photon self-energy large mu sec}).  All higher loop contributions, evaluated to leading logarithmic order, should then add up to give the full power law; since we are more interested in finite mass effects here, we will not go into the details of this calculation. 

\begin{figure}[H]
\centering
\includegraphics[width=\columnwidth]{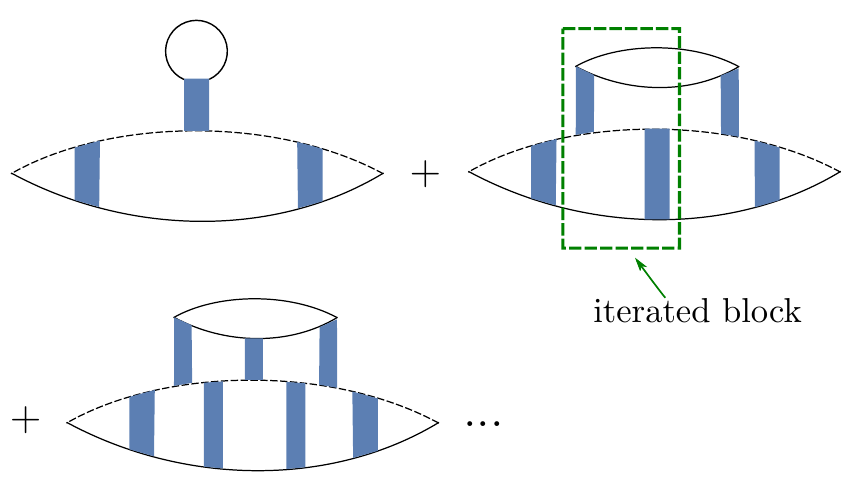}
\caption{(Color online) Series of diagrams contributing to the direct self-energy in the frequency domain. Vertical blue bars denote interaction ladders.}
\label{alldirect}
\end{figure}

To carry the diagrammatics over to finite mass, as done in the next section, it is convenient to switch to the frequency domain. A summation of all one-loop diagrams is  possible by evaluating the series shown in Fig.\ \ref{alldirect}. 

To perform the evaluation, we make use of the following simplification: 
To begin with, we often encounter complicated logarithmic integrals; however, the imaginary part of the integrand is just a delta function, so, upon integration, one finds step functions.
Since the integrand is retarded, it is then possible to recover the full expression from the imaginary part using the Kramers-Kronig relation; the step functions then become logarithms.

With that, the sum over diagrams appearing in Fig.~\ref{alldirect} assumes the form
\begin{align}
\label{Dfinalguessed}
D(\omega) &= \frac{E_B}{g^2} \frac{1}{(\omega + i 0^+)^2}  \int_{k_1<k_F} \frac{d\textbf{k}_1}{(2\pi)^2} \left\{I + I^3 + ...\right\},
\end{align}
where
\begin{align}
\label{thisisI}
I &=  \ln\left(\frac{\epsilon_{\textbf{k}_1} + \omega - \mu+i0^+}{-E_B}\right).
\end{align} 
Summing up the geometric series 
exactly reproduces the time-domain result, Eq.~(\ref{Ddirectfinal}). 
Thus, we have established how the 
photon self-energy can be calculated diagrammatically for the case of infinite VB mass $M$ (to leading order in $d_0$).


\subsection{Finite hole mass}
\label{excfinitemasssec}

We are now in a position to tackle finite VB mass $M$. Let us also consider a finite incoming momentum $\textbf{Q}$. Clearly, the one-loop criterion for choosing diagrams still holds, since we are still considering the low-density limit, $\mu \ll E_B$. We also disregard any exchange contributions for the same reasons as for the infinite mass case. As a result, we only have to recompute the series of direct diagrams of Fig \ref{alldirect}. We start with the first one. It gives:
\begin{widetext}
\begin{align}
\label{someI}
I = &-\frac{E_B V_0}{g}\!\int \displaylimits_{k_2 > k_F}\!\frac{d\textbf{k}_2}{(2\pi)^2} \frac{1}{\left(-\omega + E_B  + E(\textbf{k}_2 - \textbf{Q}) + \epsilon_{\textbf{k}_2} - \mu - i0^+\right)^2} \frac{1}{\ln\left(\frac{-E_B + \omega - \left(\textbf{Q} - \textbf{q}\right)^2/2M_\text{exc}  - \epsilon_{\textbf{k}_2} + \epsilon_{\textbf{k}_1} + i0^+}{-E_B}\right)}, 
\end{align}
\end{widetext}
where $\textbf{q} = \textbf{k}_2 - \textbf{k}_1$. The imaginary part of (\ref{someI}) reads:
\begin{align}
\label{simplified}
\nonumber
\text{Im}[I] = -\frac{V_0}{g} \int \displaylimits_{k_2 > k_F} \frac{d\textbf{k}_2}{(2\pi)^2}  \pi &\delta\left(\omega - \frac{(\textbf{Q}-\textbf{q})^2}{2M_\text{exc}} - \epsilon_{\textbf{k}_2} + \epsilon_{\textbf{k}_1}\right)  \\ & +  \mathcal{O}\left(\frac{\mu}{E_B}\right).
\end{align}
By Eq.~(\ref{simplified}), $I$ can be rewritten in a simpler form (ensuring retardation), valid for small $\omega$: 
\begin{align}
\label{datisI}
I \simeq  \frac{V_0}{g} \int\displaylimits_{k_2>k_F} \frac{d\textbf{k}_2}{(2\pi)^2} \frac{1}{\omega - \frac{(\textbf{Q} -\textbf{q} )^2}{2M_\text{exc}} - \epsilon_{\textbf{k}_2} + \epsilon_{\textbf{k}_1} + i0^+}.
\end{align}
This form can be integrated with logarithmic accuracy, which, however, only gives $\text{Re}[I]$. Specializing to $Q \ll k_F$ for simplicity, one obtains: 
\begin{align}
\label{Resimplified}
\text{Re}[I] \simeq  \ln\left(\frac{\max(|\omega + \epsilon_{\textbf{k}_1} - \mu |, \beta\mu)}{E_B}\right).
\end{align} 
As for the infinite mass case, the higher order diagrams of Fig.~\ref{alldirect} 
give higher powers of $I$. Similarly to Eq.~(\ref{Dfinalguessed}), one then obtains for the self-energy part, to leading logarithmic accuracy:
\begin{align}
\label{oneoverI}
\Sigma_\text{exc}(\textbf{Q},\omega) = -\int_{k_1 < k_F} \frac{d\textbf{k}_1}{(2\pi)^2} \cdot \frac{1}{I}.
\end{align}
The imaginary part, which determines the lineshape of $G_{\text{exc}}$, is given by
\begin{align}
\nonumber
& \text{Im}\left[\Sigma_\text{exc}(\textbf{Q},\omega)\right] \simeq  - \frac{\pi V_0}{\rho g} \int_{k_1 < k_F} \frac{d\textbf{k}_1}{(2\pi)^2}  \int_{k_2 > k_F} \frac{d\textbf{k}_2}{(2\pi)^2}   \\   & \frac{\delta(\omega - (\textbf{Q} - \textbf{q})^2/2M_\text{exc} - \epsilon_{\textbf{k}_2} + \epsilon_{\textbf{k}_1})}{\ln^2\left(\frac{\max(|\omega + \epsilon_1 - \mu|, \beta\mu)}{E_B}\right)}.
\label{ImSigmaComplicated}
\end{align}

We now apply the analogue of the logarithmic identity, Eq.~(\ref{theorem}), for a 2D integral. Thus, in leading order we may simply pull the logarithm out of the integral of Eq.~(\ref{ImSigmaComplicated}) and rewrite it as
\begin{align} 
\label{imende} \nonumber
&\text{Im}[\Sigma_\text{exc}](\textbf{Q},\omega) \simeq  -\frac{\pi V_0}{\rho g} \alpha^2 \int_{k_1 < k_F} \frac{d\textbf{k}_1}{(2\pi)^2}  \int_{k_2 > k_F} \frac{d\textbf{k}_2}{(2\pi)^2}  \\   &\qquad\qquad \delta(\omega - (\textbf{Q}-\textbf{q})^2/2M_\text{exc} - \epsilon_{\textbf{k}_2} + \epsilon_{\textbf{k}_1}).
\end{align}
The result (\ref{imende}) is physically transparent: It is just a phase-space integral giving the total rate of scattering of an exciton with momentum $\textbf{Q}$ by a CB Fermi sea electron. The prefactor is determined by the scattering phase shift $\delta$. 
At least for sufficiently small momenta $\textbf{Q}$, the integral in Eq.~(\ref{imende}) can be straightforwardly computed. For the most important case $\textbf{Q} = 0$, one obtains for small energies (see Appendix~\ref{phasespacesec}): 
\begin{align}
\label{correctnum}
\text{Im}[\Sigma_\text{exc}](\textbf{Q}=0,\omega) \sim -\alpha^2 \frac{1}{\sqrt{\beta\mu}} \theta(\omega) \omega^{3/2}, \quad \omega \ll \beta\mu,
\end{align} 
where we suppressed an irrelevant prefactor of order one. 
For $\omega \gg \beta\mu$ one recovers the infinite mass case as in (\ref{Sigmanice}).

\begin{figure}[H]
\centering
\includegraphics[width=\columnwidth]{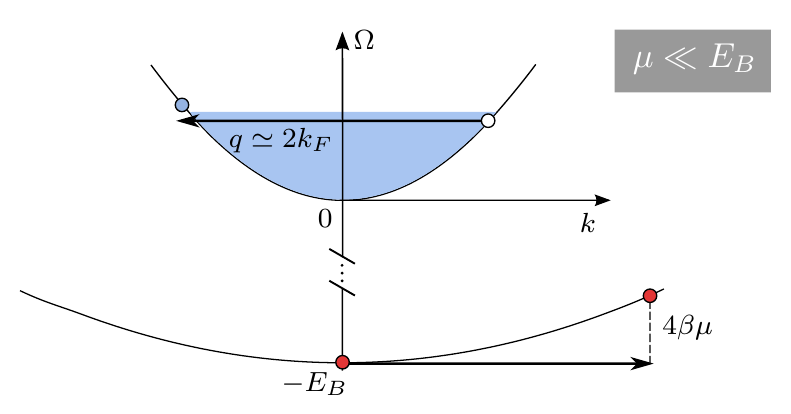}
\caption{(Color online) Scattering process of an exciton by a VB electron with large momentum transfer. The lower band represents the exciton dispersion. The scattering significantly increases the exciton energy.}
\label{recoilenergy}
\end{figure}

Compared to the infinite mass case, where $\text{Im}[\Sigma_{\text{exc}}]\sim \omega\ln(\omega)$, the self-energy (\ref{correctnum}) shows a suppression of the low-frequency scattering phase space, as seen from the higher frequency power law.
Physically, the phase space suppression is understood as follows: We have found that, after accounting for the exchange diagrams, it is admissible to view the exciton as elementary particle with mass $M_\text{exc}$, which interacts with the Fermi sea with an effective interaction strength $\alpha$ [Eq.~(\ref{alphaisphase})]. As can be seen from Fig.~\ref{recoilenergy}, scatterings of the exciton with CB electrons involving a large momentum transfer necessarily cost a finite amount of energy (the so-called recoil energy).  By contrast, in the infinite mass case such scatterings could still happen at infinitesimal energy cost, since the exciton dispersion was flat. Thus, the finite-mass phase space is reduced as compared to the infinite mass case. 
This change eventually leads to the previously asserted reappearance of the exciton delta peak.

This phase space reduction also affects the exciton spectral function, and hence the absorption: We first restrict ourselves to the leading behavior, i.e., we disregard any small renormalizations that arise from including 
 $\text{Re}[\Sigma_{\text{exc}}]$ or from higher-loop corrections. Inserting Eq.~(\ref{correctnum}) into 
Eq.\ (\ref{dressedexcwithsigma}) we then obtain, for small energies $\omega$:
\begin{align}
\label{oneoversqrt}
A(\textbf{Q} &= 0, \omega) \simeq - \Delta^2 \frac{\text{Im}[\Sigma(\omega)]}{\omega^2} \sim \Delta^2 \alpha^2 \frac{\theta(\omega)}{\sqrt{\beta\mu\cdot \omega}},
\end{align}
with
\begin{align}
 \Delta^2 &= \frac{d_0^2\rho E_B}{g^2}. \label{Deltadef}
\end{align}   
The factor $\Delta$ (with units of energy) determines the polariton splitting at zero detuning, and will be discussed in Sec.~\ref{Polariton properites sec}.
The $1/\sqrt{\omega}$ divergence seen in (\ref{oneoversqrt}) was also found by Rosch and Kopp using a path-integral approach \cite{PhysRevLett.75.1988} for a related problem, that of a heavy hole propagating in a Fermi sea. In addition, Rosch and Kopp find a quasi particle delta peak with a finite weight. This peak can also be recovered within our approach upon inclusion of the correct form of $\text{Re}[\Sigma_{\text{exc}}]$.
From Eqs.~(\ref{Resimplified}) and (\ref{oneoverI}) we may infer it to be 
\begin{align}
\label{Reinfer}
\text{Re}[\Sigma_{\text{exc}}(\textbf{Q}=0,\omega)] = \alpha^2 \omega \ln\left(\frac{\sqrt{\omega^2 + (\beta\mu)^2}}{\mu}\right),
\end{align}
where we have rewritten the maximum-function with logarithmic accuracy using a square root. 
This cut-off of logarithmic singularities (which are responsible for edge power laws) by recoil effects is a generic feature of our model,  
and will reoccur in the regime of $\mu \gg E_B$ presented in 
Sec.~\ref{Photon self-energy large mu sec}. In qualitative terms, this is also discussed in Ref.\ \cite{Nozi`eres1994} (for arbitrary dimensions).
Our results are in full agreement with this work.

We may now deduce the full photon self-energy $\Pi_{\text{exc}}$ as follows: In the full finite-mass version of the power law (\ref{Nozieresresult}), the real part of the logarithm in the exponent will be replaced by the cut-off logarithm from Eq.~(\ref{Reinfer}). The imaginary part of this logarithm will be some function $f(\omega)$ which continuously interpolates between the finite-mass regime for $\omega \ll \beta \mu$ [given by Eq.~(\ref{correctnum}) times $\omega^{-1}$], and the infinite mass regime for $\omega \gg \beta\mu$. 
Therefore, we arrive at
\begin{align}
\label{Piexcfinitemass}
&\Pi_{\text{exc}}(\textbf{Q} = 0,\omega) =  \\ &\frac{\Delta^2}{\omega+i0^+} \exp \left[\alpha^2 \left(\ln\left(\frac{\sqrt{\omega^2 + (\beta\mu)^2}}{\mu}\right) - if(\omega)\right)\right]   \nonumber,
\end{align}
where
\begin{align}
&f(\omega) =
\begin{cases}
 \pi \sqrt{\frac{\omega}{\beta\mu}} \theta(\omega)  \quad &\omega \ll \beta\mu \\ 
\pi  \quad &\omega \gg \beta\mu.
\end{cases}
\end{align}
It is seen by direct inspection that (\ref{Piexcfinitemass}) has a delta peak at $\omega =0$ with weight $\Delta^2 \beta^{\alpha^2}$.

One can also asses the weight of the delta peak by comparing the spectral weights of the exciton spectral function in the infinite and finite mass cases: 
The weight of the delta peak must correspond to the difference in spectral weight as the absorption frequency power law is changed
once $\beta$ becomes finite.
In the infinite mass case, the absorption scales as 
\begin{align}
A_\infty(\omega)\sim\frac{\Delta^2 \alpha^2}{\omega}  \left(\frac{\omega}{\mu}\right)^{\alpha^2}\theta(\omega),
\end{align}
as follows from Eq.~(\ref{Nozieresresult}) above. 
Thus, the spectral weight in the relevant energy region is given by 
\begin{align} 
\label{masspoleweight}
 \int_0^{\beta\mu} d\omega A_{\infty}(\omega) = \Delta^2 \beta^{{\alpha}^2}.
\end{align}
In contrast, using Eq.~(\ref{correctnum}), the spectral weight of the finite mass case is 
\begin{align}
 \int_0^{\beta\mu} d\omega A(\textbf{Q}=0,\omega) = \Delta^2 \alpha^2.
\end{align}
For scattering phase shifts $\delta$ close to $\pi$ (i.e., $\alpha \rightarrow 0$), and for finite mass, $\beta>0$, a pole with weight proportional to $\beta^{\alpha^2}$ [Eq.~(\ref{masspoleweight})] at $\omega =0$ should be present in the spectrum, if $\beta$ is not exponentially small in $\alpha$.
This weight is exactly the same as for the heavy hole when computed in a second order cumulant expansion~\cite{PhysRevLett.75.1988}.

The full imaginary part of $\Pi_\text{exc}(\textbf{Q}=0,\omega)$ was already given explicitly in Eqs.~(\ref{Excgeneral}) and (\ref{Exccases}), and plotted in Fig.~\ref{finmasssmallmu1}.
That plot illustrates the main conclusion of this section: For finite mass, Fermi sea excitations with large momentum transfer are energetically unfavorable, and are therefore absent from the absorption power law. As a result, the pole-like features of the absorption are recovered.

\subsection{Validity of the electron-hole correlator as a photon self-energy}
\begin{figure}[H]
\centering
\includegraphics[width=\columnwidth]{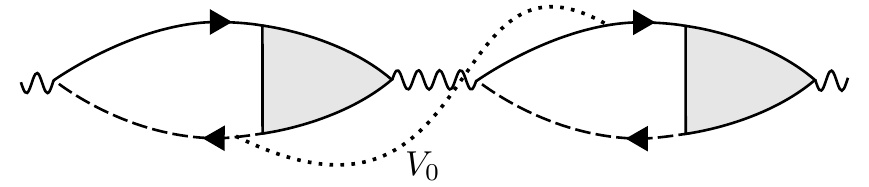}
\caption{Two dressed bubbles, connected by one electron-hole interaction (dotted line). 
This is an example of a photon self-energy diagram that is not contained in our approximation for $\Pi(\textbf{Q},\omega)$.}
\label{twobubbles}
\end{figure}

Let us now assess the validity of the expressions for the CB electron-VB hole correlator [Eqs.~(\ref{Nozieresresult}) and (\ref{Piexcfinitemass})] as a photon self-energy. Using them, one assumes that only electron-hole interactions within one bubble are of relevance, and electron-hole interactions connecting two bubbles (an example is shown in Fig.~\ref{twobubbles}) can be disregarded.

The regime where such an approximation is valid may be inferred from the following physical argument: 
Electronic processes (i.e. electron-hole interactions) happen on the time scale of Fermi time $1/\mu$. On the other hand, the time scale for the emission and reabsorption of a photon (which is the process separating two bubbles) is given by $1/\rho d_0^2$ (where $d_0$ is the dipole matrix element). If the second scale is much larger than the first one, electrons and holes in distinct bubbles do not interact. Thus, the our approach is valid as long as
\begin{align}
\label{M0smallerthanmu}
\rho d_0^2 \ll \mu.
\end{align}
Under this condition, the following physical picture is applicable: an exciton interacts with the Fermi sea, giving rise to a broadened exciton, which in turn couples to the cavity photons. When Eq.~(\ref{M0smallerthanmu}) is violated, one should think in different terms: excitons couple to photons, leading to exciton-polaritons. These then interact with the Fermi sea. The second scenario is, however, beyond the scope of this paper. 

The above discussion is likewise valid for the regime of large Fermi energy, which is studied below.

\section{Electron-hole correlator for large Fermi energy}
\label{Photon self-energy large mu sec}

We now switch to the opposite regime, where $\mu \gg E_B$, and excitons are not well-defined.
For simplicity, we also assume that $\mu$ is of the order of the CB bandwidth. 
Hence, $E_B \ll \mu \simeq \xi$.
Within our simplified model, the finite mass problem in 3D was solved in \cite{gavoret1969optical}. This treatment can be straightforwardly carried over to 2D \cite{Pimenov2015}. To avoid technicalities, we will, however, just show how to obtain the 2D results in a ``Mahan guess'' approach~\cite{PhysRev.163.612}, matching known results from~\cite{PhysRevB.35.7551}. To this end, we will first recapitulate the main ingredients of the infinite mass solution.

\subsection{Infinite hole mass}
The FES builds up at the Burstein-Moss shifted threshold $\Omega_T^{\text{FES}} = E_G + \mu$.
Its diagrammatic derivation relies on a weak-coupling ansatz: The parameter $g = \rho V_0$ is assumed to be small. As seen from Eq.~(\ref{gestimate}), this is indeed true for $\mu \gg E_0$. 
In principle, below the FES  there will still be the exciton peak; however, this peak will be broadened into a weak power law, and thus merge with the FES. For finite mass (see below), the position of the would-be exciton may even be inside FES continuum, which makes the exciton disappear completely. What is more, the exciton weight, being proportional to $E_B$, is exponentially small in $g$ (since $\mu \simeq \xi$). We may therefore safely disregard the exciton altogether (see also discussion in Appendix \ref{muincapp}).
  
\begin{figure}[H]
\centering
\includegraphics[width=.6\columnwidth]{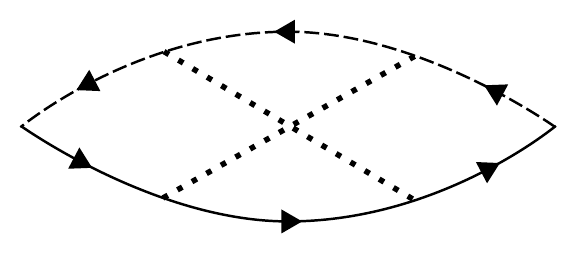}
\caption{Lowest order crossed diagram contributing to the FES.}
\label{crossed_infmass}
\end{figure}

To leading order in $g\ln(\omega/\mu)$, the dominant contribution comes from the so called ``parquet'' diagrams, containing all possible combinations of ladder and crossed diagrams~\cite{PhysRev.178.1072, NOZIERES1969}. 
The value of the pure ladder diagrams is given by Eq.~(\ref{ladder contribution}), with $\Omega - E_G$ replaced by $\omega = \Omega -\Omega_T^{\text{FES}}$.
The lowest-order crossed diagram is shown in Fig.~\ref{crossed_infmass}.
With logarithmic accuracy the contribution of this diagram is easily computed: 
\begin{align}
\Pi_{\text{crossed}}
=  -\frac{1}{3}d_0^2\rho g^2 \left[\ln(\omega/\mu)\right]^3.
\end{align}
This is $-1/3$ times the contribution of the second order ladder diagram, c.f.\ Eq.~(\ref{ladder contribution}). Thus, the ladder and crossed channels partially cancel each other, a feature which persists to all orders. This also shows that the FES is qualitatively different from the broadened exciton discussed in the previous section: now the exciton effects (ladder diagrams) and the Fermi sea shakeup (crossed diagrams) have to be treated on equal footing.

In his original paper Mahan computed all leading diagrams to third order and guessed the full series from an exponential ansatz~\cite{PhysRev.163.612}. The corresponding result for the photon self-energy $\Pi_{\text{FES}}(\omega)$ reads
\begin{align}
\label{Mahanresult}
\Pi_{\text{FES}}(\omega) = \frac{d_0^2\rho}{2g}\left(1-\exp\left[-2g\ln\left(\frac{\omega+i0^+}{-\mu}\right)\right]\right).
\end{align}
Relying on coupled Bethe-Salpeter equations in the two channels (ladder and crossed), Nozi\`{e}res \textit{et al.}\ then summed all parquet diagrams, where a bare vertex is replaced by (anti-)parallel bubbles any number of times~\cite{PhysRev.178.1072, NOZIERES1969}. The result corresponds exactly to  Mahan's conjecture, Eq.~(\ref{Mahanresult}).
  
By the standard FES identification 
$\delta/\pi = g + \mathcal{O}(g^3)$, the power law in Eq.~(\ref{Mahanresult}) coincides with the one given in Eq.~(\ref{Nozieresresult}); the phase shift is now small. 
One should also point out that the peaks in the spectra in the regimes of small $\mu$ (Fig.~\ref{finmasssmallmu1}) and large $\mu$ (Fig.~\ref{FEScomp}) are not continuously connected, since the FES arises from the continuous threshold, whereas the exciton does not. 

Let us finally note that since $\mu$ is a large scale, 
Eq.~(\ref{Mahanresult}) should be a good approximation for the 
photon self-energy, since the condition (\ref{M0smallerthanmu}) is easily satisfied.

\subsection{Finite hole mass}
\label{FESfiniteholemasssubseq}
As in the regime of the exciton, in the finite mass case the result~(\ref{Mahanresult}) will be modified due to the recoil energy $\beta\mu$. However, it will now be the \textit{VB hole} recoil (or the hole lifetime, see below) instead of the exciton recoil ---  the latter is meaningless since the exciton is not a well defined entity anymore. This is most crucial: 
Since CB states with momenta smaller than $k_F$ are occupied, VB holes created by the absorption of zero-momentum photons must have momenta larger than $k_F$. Therefore, the hole energy can actually be lowered by scatterings with the Fermi sea that change the hole momenta to some smaller value, and these scattering processes will cut off the sharp features of $\Pi_{\text{FES}} (\omega)$. 
The actual computation of the photon self-energy with zero photon momentum, $\Pi_{\text{FES}}(\textbf{Q}=0,\omega)$, proceeds in complete analogy to the 3D treatment of~\cite{gavoret1969optical}. Limiting ourselves to the ``Mahan guess'' for simplicity, the main steps are as follows.

The first major modification is the appearance of two thresholds: As easily seen by the calculation of the ladder diagrams, the finite mass entails a  
shift of the pole of the logarithm from $\omega =0$ to $\omega = \beta\mu$, which is the minimal energy for direct transitions obeying the Pauli principle. Correspondingly, $\omega_D=\beta\mu$ is called the direct threshold.  Near this threshold, logarithmic terms can be large, and a non-perturbative resummation of diagrams is required. However, the true onset of 2DEG absorption will actually be the indirect threshold $\omega_I=0$. There, the valence band hole will have zero momentum, which is compensated
by a low-energy conduction electron-hole pair, whose net momentum is $-k_F$. The two thresholds were shown in Fig.~\ref{twothresholds}.
It should be noted that for $E_B < \beta\mu$
the exciton energy $\approx \omega_D - E_B$, 
is between $\omega_I$ and $\omega_D$. Hence, in this case the exciton overlaps with the continuum and is completely lost.

\begin{figure}[H]
\centering
\includegraphics[width=.6\columnwidth]{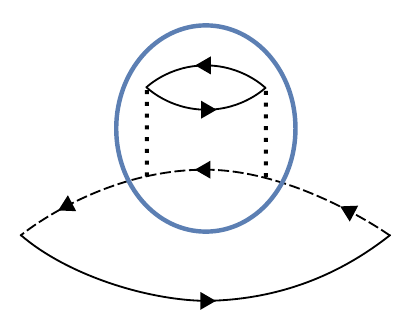}
\caption{(Color online) Second diagram (in addition to Fig.~\ref{crossed_infmass}) contributing to the absorption at the indirect threshold $\omega_I$. The blue ellipse marks the VB self-energy insertion used below.}
\label{omega3self}
\end{figure}

Near $\omega_I$, the problem is completely perturbative. In leading (quadratic) order in $g$, the absorption is determined by two diagrams only. The first one is the crossed diagram of Fig.~\ref{crossed_infmass}. The second one is shown in Fig.~\ref{omega3self}.
When summing these two diagrams, one should take into account spin, which will simply multiply the diagram of Fig.~\ref{omega3self} by a factor of two (if the spin is disregarded, the diagrams will cancel in leading order). Up to prefactors of order one, the phase-space restrictions then result in a 2DEG absorption~(see \cite{PhysRevB.35.7551} and Appendix~\ref{phasespacesec}): 
\begin{align}
\label{Abspowerlaw}
A(\textbf{Q} = 0,\omega) = d_0^2 g^2 \left(\frac{\omega}{\beta\mu}\right)^3 \theta(\omega).
\end{align}
The phase space power law $\omega^3$ is specific to 2D . Its 3D counterpart has a larger exponent, $\omega^{7/2}$ \cite{PhysRevB.35.7551}, due to an additional restriction of an angular integration.

\begin{figure}[H]
\centering
\includegraphics[width=.5\columnwidth]{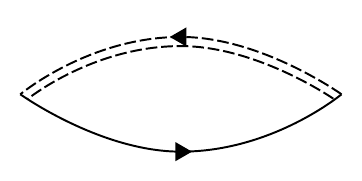}
\caption{The CB electron-VB hole bubble, with the hole propagator dressed by the self-energy, Eq.~(\ref{imselfhole}).}
\label{dressedbubble}
\end{figure}

Let us now turn to the vicinity of
$\omega_D$, where one has to take into account the logarithmic singularities and the finite hole life-time in a consistent fashion. Regarding the latter, 
one can dress all VB lines with self-energy diagrams as shown in Fig.~\ref{omega3self}. The self-energy insertion at the dominant momentum $k = k_F$ reads
\begin{align}
\label{VBself-energy}
\text{Im}[\Sigma_{\rm{VB}}(k_F, \omega)] = \frac{1}{\sqrt{3}}\theta(\omega)  g^2 \beta\mu \frac{\omega^2}{(\beta\mu)^2}, \quad \omega \ll \beta\mu.
\end{align} 
As can be shown by numerical integration, this expression reproduces the correct order of magnitude for $\omega = \beta\mu$, such that it can be safely used in the entire interesting regime $\omega \in [0,\beta\mu]$. The power law in Eq.~(\ref{VBself-energy}) is again specific to 2D. In contrast, the order of magnitude of the inverse lifetime is universal,
\begin{align}
\label{imselfhole}
 \text{Im}[\Sigma_{\rm{VB}}(k_F, \beta\mu)] \sim g^2\beta\mu.
\end{align}
Disregarding the pole shift arising from $\text{Re}[\Sigma]$, the self-energy (\ref{imselfhole}) can be used to compute the ``dressed bubble'' shown in Fig.~\ref{dressedbubble}.
With logarithmic accuracy, the dressed bubble can be evaluated analytically. In particular, its real part  
reads:
\begin{align}
\label{relogdressed}
\text{Re}\left[\Pi_{\text{db}}\right](\omega) \simeq 
{\rho d_0^2}\ln\left(\frac{\sqrt{(\omega - \beta\mu)^2 + \left(g^2 \beta\mu \right)^2}}{\mu}\right).
\end{align}
This is just a logarithm whose low-energy divergence is cut by the VB hole life time, in full analogy to Eq.~(\ref{Reinfer}),  and in agreement with Ref.\ \cite{Nozi`eres1994}.

\begin{figure}[H]
\centering
\includegraphics[width=\columnwidth]{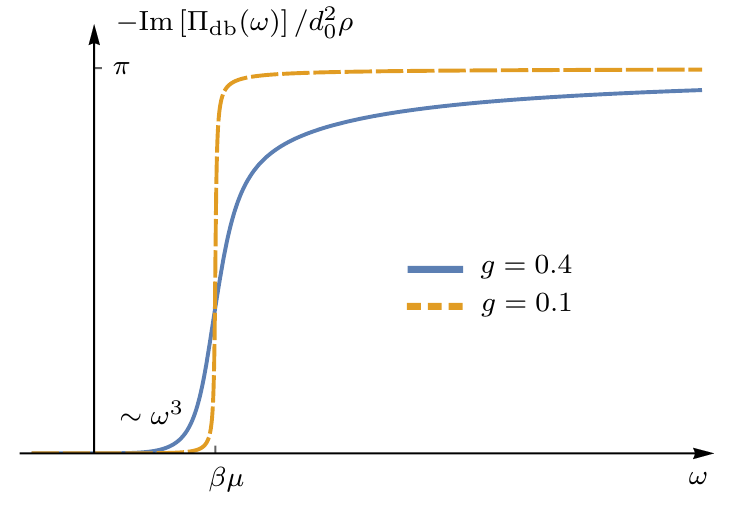}
\captionsetup[justification]{justified}
\caption{(Color online) Imaginary part of the dressed bubble for two values of $g$, obtained from numerical integration of $\Pi_{
\text{db}}$, using the hole self-energy insertion of (\ref{VBself-energy}).}
\label{Imdressedbubbleplot}
\end{figure}

For the computation of polariton spectra later on, it  turns out to be more practical to obtain both the real and the imaginary parts of $\Pi_{\text{db}}(\omega)$ by numerically integrating the approximate form \cite{Pimenov2015}:

\begin{align}
\label{contclosed}
&\Pi_{\text{db}}(\omega) \simeq \\&
\notag 
\frac{d_0^2}{(2\pi)^2}\hspace{-.8em}\int\displaylimits_{k > k_F}\hspace{-.8em}d\textbf{k} \frac{1}{\omega - (\epsilon_{\textbf{k}}-\mu) - \frac{k^2}{2M} +  i\text{Im}[\tilde{\Sigma}_{\rm{VB}}(\omega - \epsilon_\textbf{k} + \mu)]}, \\& \notag
\text{Im}[\tilde{\Sigma}_{\rm{VB}}(x)]= \begin{cases}  \tfrac{g^2}{\sqrt{3}}\theta(x)  \frac{x^2}{(\beta\mu)}  & x< 
\beta\mu\\
\tfrac{g^2}{\sqrt{3}}\beta\mu & x>\beta\mu,
\end{cases}
\end{align} 
to avoid unphysical spikes arising from the leading logarithmic approximation. A corresponding  plot of $-\text{Im}\left[\Pi_{\text{db}}\right]$ is shown in Fig.~\ref{Imdressedbubbleplot}.
The numerical expression $-\text{Im}\left[\Pi_{\text{db}}\right]$ simplifies to the correct power law (\ref{Abspowerlaw}) in the limit $\omega \rightarrow 0$, and  approaches the infinite mass value $  d_0^2\rho\pi$ for large frequencies.

Higher-order diagrams will contain higher powers of the rounded logarithm~(\ref{relogdressed}). The parameter controlling the leading log scheme now reads 
\begin{align}
l\equiv g\ln(\beta g^2).
\end{align}
One can distinguish different regimes of $l$.
The simplest is
$l \ll 1$, which holds in the limit $g \rightarrow 0$ (or, put differently, if $\beta$ is not exponentially small in $g$). In this limit, no singularity is left. The large value of the Fermi energy (small $g$) and the large value of the hole decay $\beta\mu$ have completely overcome all interaction-induced excitonic effects. A decent approximation to the 2-DEG absorption is then already given by the imaginary part of the dressed bubble. Fig.~\ref{Imdressedbubbleplot} shows the corresponding absorption. 

The more interesting regime corresponds to $g\ln(\beta  g^2) \gtrsim 1$, where arbitrary numbers of conduction band excitations contribute to the absorption alike 
 \footnote{The regime of $g\ln(\beta g^2) \gg 1$ is out of reach for the methods used in~\cite{gavoret1969optical}. To study it, a consistent treatment of the  divergences is needed, similar to~\cite{NOZIERES1969}. We will not attempt this here}. A non-perturbative summation is needed, which is, however, obstructed by the following fact:
As found by straightforward computation, the crossed diagrams are not only cut by $g^2\beta\mu$ due to the hole decay, but also acquire an inherent cutoff of order $\beta\mu$ due to the hole recoil. A standard parquet summation is only possible in a regime where these two cutoffs cannot be distinguished with logarithmic accuracy, i.e.\ where $\beta \ll g^2$. For small enough $g$ this will, however, always be the case in the truly non-perturbative regime where $\beta$ must be exponentially small in $g$.

As a result of these considerations, the logarithms of the parquet summation have to be replaced by the cut-off logarithms~(\ref{relogdressed}), with $g^2\beta\mu$ replaced by $\beta\mu$. The imaginary part of the logarithm is then given by the function plotted in Fig.~\ref{Imdressedbubbleplot}.
The resulting full photon self-energy in the non-perturbative FES regime reads:
\begin{align}
\label{Pillow}
\Pi_\text{FES}(\textbf{Q}=0,\omega) &\simeq -\frac{d_0^2\rho}{2g}\left(\exp\left[-2g\left(\frac{\Pi_{\text{db}}(\omega)}{\rho d_0^2}\right)\right] -1\right).
\end{align}
A sketch of $\text{Im} \left[\Pi_{\text{FES}}\right]$ is shown in Fig.~\ref{FEScomp}.

\section{Polariton properties}
\label{Polariton properites sec}
When the cavity energy $\omega_c$ is tuned into resonance with the excitonic 2DEG transitions, the matter and light modes hybridize, resulting in two polariton branches. We will now explore their properties in the different regimes.

\subsection{Empty conduction band}
\begin{figure}[H] 
\centering
\includegraphics[width=\columnwidth]{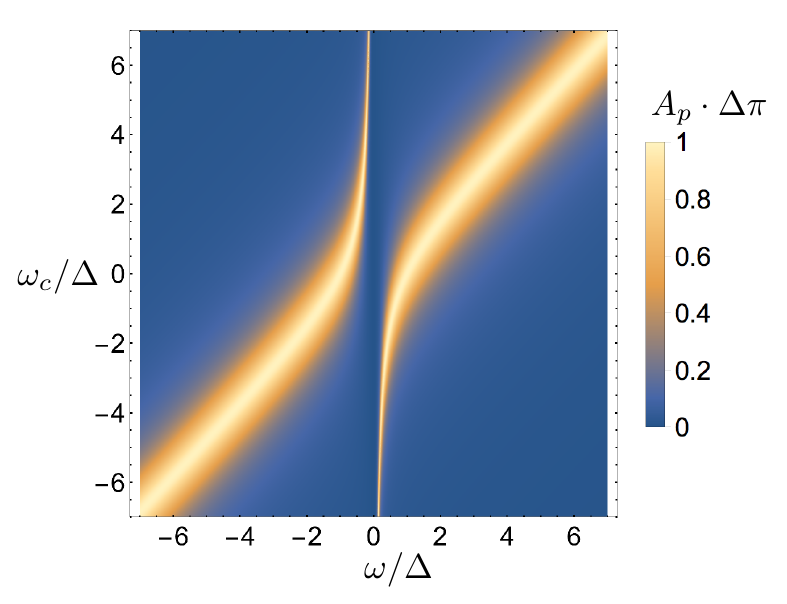}
\caption{(Color online) $\mu = 0$: Exciton-polariton spectrum as function of cavity detuning $\omega_c$ and energy $\omega$, measured in units of the half polariton splitting $\Delta$, with 
$\Gamma_c = \Delta$. 
 }
\label{pureexcitonpolariton}
\end{figure}

To gain some intuition, it is first useful to recapitulate the properties of the exciton-polariton in the absence of a Fermi sea. Its (exact) Green's function is given by Eq.~(\ref{dressedphot}), with 
$\omega_\textbf{Q=0} = \omega_c$ and $\Pi(\omega) = \Delta^2/{(\omega+i0^+)}
$, where $\Delta$ is a constant (with units of energy) which determines the polariton splitting at zero detuning. In terms of our exciton model, one has $\Delta =  \sqrt{d_0^2\rho E_B/g^2}$. $\omega$ is measured from the exciton pole. 
A typical density plot of the polariton spectrum $A_p = -\text{Im}\left[D^R(\omega,\omega_c)\right]/\pi$, corresponding to optical (absorption) measurements as e.g.\ found in \cite{Smolka}, is shown in Fig.\ \ref{pureexcitonpolariton}. 
A finite cavity photon linewidth $\Gamma_c = \Delta$ is used. 
The physical picture is transparent: the bare excitonic mode (corresponding to the vertical line) and the bare photonic mode repell each other, resulting in a symmetric avoided crossing of two polariton modes. 

For analytical evaluations, it is more transparent to consider an infinitesimal cavity linewidth $\Gamma_c$. The lower and upper polaritons will then appear as delta peaks in the polariton spectral function, at positions
\begin{align}
\omega_\pm = \frac{1}{2} \left(\omega_c \pm \sqrt{\omega_c^2 + 4\Delta^2}\right),
\end{align}
and with weights 
\begin{align}
\label{Weightsexcitonexact}
W_\pm = \frac{1}{1 + \frac{4 \Delta^2}{(\omega_c \pm \sqrt{4 \Delta^2 + \omega_c^2})^2}}.
\end{align}
We note that the maximum of the polariton spectra scales as $1/\Gamma_c$ for finite $\Gamma_c$.
Our spectral functions are normalized such that the total weight is unity. From Eq.~(\ref{Weightsexcitonexact}) it is seen that the weight of the ``excitonic'' polaritons (corresponding to the narrow branches of Fig.~\ref{pureexcitonpolariton}) decays as $\Delta^2/\omega_c^2$ for large absolute values of $\omega_c$.

\subsection{Large Fermi energy}
Let us study polariton properties in the presence of a Fermi sea. Reverting the order of presentation previously taken in the paper, we first turn to the regime of large Fermi energy, $E_B \ll \mu$. 
This is because for $E_B \ll \mu$ the inequality $\rho d_0^2 \ll \mu$~(\ref{M0smallerthanmu}) is more easily satisfied than in the opposite limit of $E_B \gg \mu$, facilitating experimental realization. We  compute the polariton properties using the electron-hole correlators as cavity photon self-energy.
 A similar approach was applied recently by Averkiev and Glazov~\cite{PhysRevB.76.045320}, who computed cavity transmission coefficients semiclassically, phenomenologically absorbing the effect of the Fermi-edge singularity into the dipole matrix element. Two further recent treatments of polaritons for nonvanishing Fermi energies are found in \cite{PhysRevB.89.245301} and \cite{baeten2015mahan}. In the first numerical paper \cite{PhysRevB.89.245301}, the Fermi-edge singularity as well as the excitonic bound state are accounted for, computing the electron-hole correlator as in~\cite{combescot1971infrared}, but an infinite mass is assumed. The second paper~\cite{baeten2015mahan} is concerned with finite mass. However, the authors only use the ladder approximation and neglect the crossed diagrams, partially disregarding the physical ingredients responsible for the appearance of the Fermi-edge power laws. We aim here to bridge these gaps and describe the complete picture in the regime of large Fermi energy (before turning to the opposite regime of $\mu \ll E_B$).

In the infinite mass limit we will use Eq.~(\ref{Mahanresult}) as the photon self-energy. It is helpful to explicitly write down the real and imaginary parts of the self-energy in leading order in $g$: 
\begin{align}
\label{ReFermiInfinite}
\text{Re}\left[\Pi_{\text{FES}}\right](\omega) &= \tilde{\Delta} \left(1-\left(\frac{|\omega|}{\mu }\right)^{-2g}\right),\\
\label{AbsFermiInfinite}
\text{Im}\left[\Pi_{\text{FES}}\right](\omega) &= - \tilde{\Delta} \cdot 2\pi g\left(\frac{\omega}{\mu }\right)^{-2g}\theta(\omega) \\
\label{deltatildedef}
\tilde{\Delta} &\equiv \frac{d_0^2\rho}{2g},
\end{align}
where we have introduced the parameter $\tilde{\Delta}$, which determines the splitting of the polaritons, playing a similar role to $\Delta$ in the previous case of empty CB. In the following, $\tilde{\Delta}$ will serve as the unit of energy.

\begin{figure}[H] 
\centering
\includegraphics[width=\columnwidth]{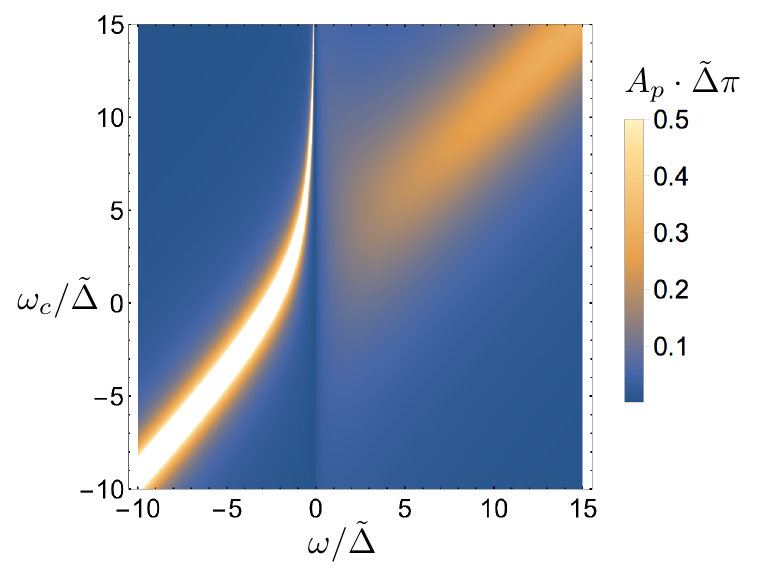}
\caption{(Color online) $\mu \gg E_B$: Infinite hole mass Fermi-edge-polariton spectrum $A_p(\omega,\omega_c)$ as function of cavity detuning $\omega_c$ and energy $\omega$, measured in units of the effective splitting $\tilde{\Delta}$. It was obtained by inserting Eqs.~(\ref{ReFermiInfinite}) and~(\ref{AbsFermiInfinite}) into Eq.~(\ref{Polaritonspectralfunction}). Parameter values: $\mu  = 30\tilde{\Delta}$, $\Gamma_c = 1\tilde{\Delta}$, and $g=0.25$.}
\label{Fermipolaritoninfinitemass}
\end{figure}

For a cavity linewidth $\Gamma_c = 1\tilde{\Delta}$, a typical spectral plot of the corresponding "Fermi-edge polaritons" is shown in Fig.~\ref{Fermipolaritoninfinitemass}. It is qualitatively similar to the results of~\cite{PhysRevB.76.045320}.
A quantitative comparison to the empty CB case is obviously not meaningful due to the appearance of the additional parameters $\mu$ (units of energy) and $g$ (dimensionless). Qualitatively, one may say the following: The lower polariton is still a well-defined spectral feature. For zero cavity linewidth (see below), its lifetime is infinite. The upper polariton, however, is sensitive to the high-energy tail of the 2DEG absorption power law~(\ref{AbsFermiInfinite}), and can decay into the continuum of CB particle-hole excitations. Its linewidth is therefore strongly broadened. Only when the 2DEG absorption is cut off by finite bandwidth effects (i.e., away from the Fermi-edge), a photonic-like mode reappears in the spectrum (seen in the upper right corner of Fig.~\ref{Fermipolaritoninfinitemass}). 

\begin{figure}[H]
\centering
\includegraphics[width=\columnwidth]{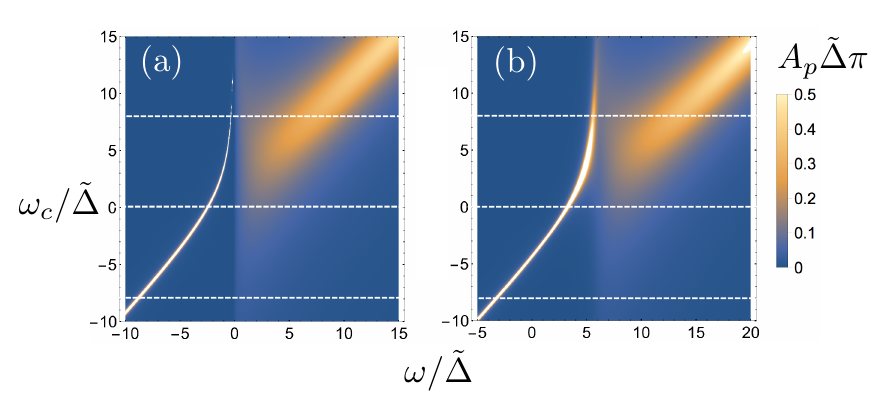}
\caption{(Color online) $\mu \gg E_B$: 
(a) Fermi-edge-polariton spectrum with the same parameters as in Fig.~\ref{Fermipolaritoninfinitemass}, but $\Gamma_c=0.01\tilde{\Delta}$. The white dashed lines denote the location of the spectral cuts presented in Fig.~\ref{combined_cuts}.
(b) Spectrum with a nonzero mass-ratio $\beta = 0.2$, and otherwise the same parameters as in (a). This plot was obtained by inserting the finite mass photon self-energy of Eq.~(\ref{Pillow}) into Eq.~(\ref{Polaritonspectralfunction}), with $\omega_c$ replaced by $\omega_c + \beta\mu$ to make sure that the cavity detuning is measured from the \textit{pole} of the photon self-energy. Note that the frequency range of panel~(b) is shifted as compared to~(a).}
\label{combined_spectra}
\end{figure}

For more detailed statements, one can again consider the case of vanishing cavity linewidth $\Gamma_c$. A spectral plot with the same parameters as in Fig.~\ref{Fermipolaritoninfinitemass}, but with small cavity linewidth, $\Gamma_c = 0.01 \tilde{\Delta}$, is shown in Fig.~\ref{combined_spectra}(a).

\begin{figure*}[!]
\centering
\includegraphics[width=\textwidth]{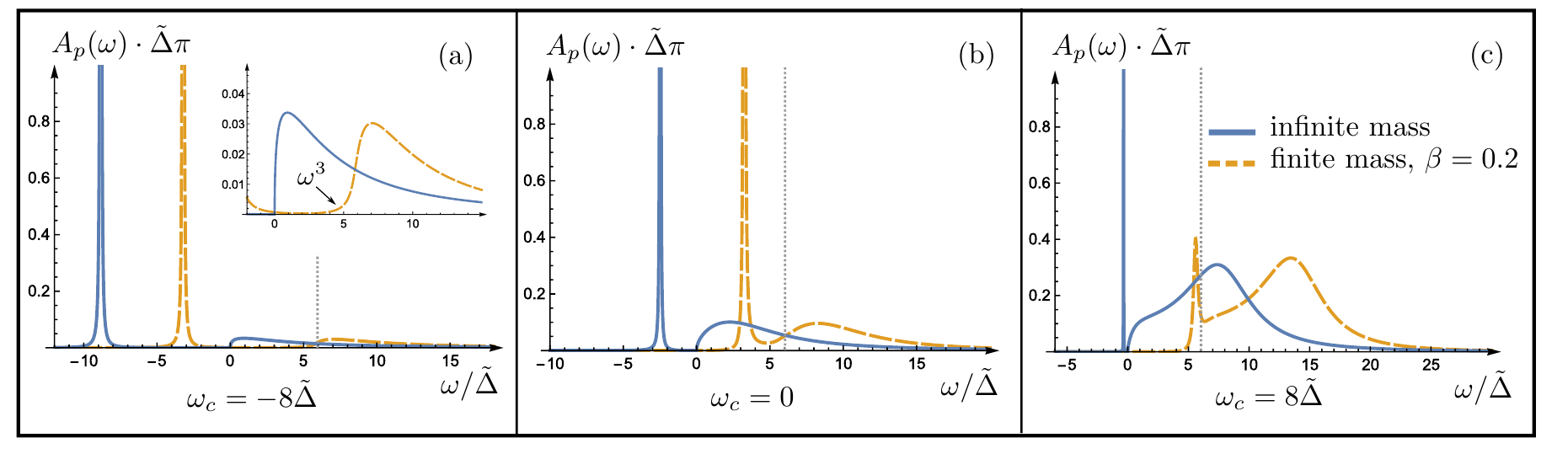}
\caption{(Color online) $\mu \gg E_B$: Spectral cuts at fixed cavity detuning through the polariton spectra of  Fig.~\ref{combined_spectra}, for both infinite (continuous blue lines) and finite (dashed orange lines) hole mass. 
(a) Large negative cavity detuning. The dotted vertical line line always indicates the position of the direct threshold at $\omega = \beta\mu$. The inset 
is a zoom-in on the absorption onset at the indirect threshold.
(b) Zero cavity detuning. 
(c) Large positive cavity detuning. 
}
\label{combined_cuts}
\end{figure*}

We first examine the lower polariton (assuming zero linewidth), which is a pure delta peak. Its position is determined by the requirement
\begin{align}
\label{findlowerpole}
\omega-\omega_c - \text{Re}\left[\Pi_\text{FES}(\omega)\right] = 0.
\end{align}
One may study the solution of this equation in three distinct regimes, corresponding to $\omega_c \rightarrow -\infty$, $\omega_c = 0$, and $\omega_c \rightarrow + \infty$.

For $\omega_c \rightarrow - \infty$, the solution of Eq.~(\ref{findlowerpole}) approaches $\omega = \omega_c$, and the lower polariton acquires the full spectral weight (unity): For strong negative cavity detunings, the bare cavity mode is probed. The corresponding spectral cut is shown in Fig.~\ref{combined_cuts}(a) (continuous line). We will refrain from making detailed statements about the way the bare cavity mode is approached, since this would require the knowledge of the photon self-energy at frequencies far away from the threshold.
As the cavity detuning is decreased, the lower polariton gets more matter-like. At zero detuning [see Fig.~\ref{combined_cuts}(b)], and for $g$ not too small (w.r.t.\ $g\tilde{\Delta}/\mu$), the weight of the lower polariton is approximately given by $1/(1+2g)$.
For large positive cavity detunings [see Fig.~\ref{combined_cuts}(c)], the position of the matter-like lower polariton approaches $\omega=0$,
\begin{align} 
\label{peaklargewc}
\omega \sim -\omega_c^{-1/(2g)} \quad \text{as} \quad \omega_c \rightarrow \infty. 
\end{align} 
The lower polariton weight also scales in a power law fashion, 
$\sim \omega_c^{-1-1/(2g)}$, distinct from the excitonic regime, where the weight falls off quadratically [Eq.~(\ref{Weightsexcitonexact})].

Due to the finite imaginary part of the self-energy $\Pi_{\text{FES}}(\omega)$, the upper polariton is much broader than the lower one: the photonic mode can decay into the continuum of matter excitations. At large negative detunings [see the inset to Fig.~\ref{combined_cuts}(a)], the upper polariton has a power law like shape (with the same exponent as the Fermi-edge singularity), and for $
\omega_c \rightarrow - \infty$ its maximum approaches $\omega = 0$ from the high-energy side. As the detuning is increased (made less negative), the maximum shifts away from $\omega=0$, approaching the free cavity mode frequency $\omega = \omega_c$ for $\omega_c \rightarrow \infty$. Since the weight and height are determined by the value of $\text{Im}[\Pi_{\text{FES}}]$ at the maximum, they increase correspondingly.

Let us now consider the case of finite mass. Using the finite mass photon self-energy (\ref{Mahanresult}) instead of (\ref{Pillow}), the Fermi-edge-polariton spectrum with a  nonzero mass-ratio of $\beta = 0.2$ is plotted in Fig.~\ref{combined_spectra}(b).
Compared to the infinite mass case of Fig.~\ref{combined_spectra}(a), Fig.~\ref{combined_spectra}(b) has the following important features: 
(i) The boundary line separating the lower and upper thresholds is shifted to the high-energy side from $\omega = 0$ in the infinite mass case to $\omega = \beta\mu$ in the finite mass case, reflecting the Burstein-Moss shift in the 2DEG absorption.
(ii) As opposed to the infinite mass case, the lower polariton is strongly broadened at large positive detunings.

These points are borne out more clearly in Fig.~\ref{combined_cuts}(a)--(c) (dashed lines), which presents cuts through Fig.~\ref{combined_spectra}(b) at fixed detuning.
The situation at large negative detuning is shown in Fig.~\ref{combined_cuts}(a): Compared to the infinite mass case, shown as full line, the polaritons are shifted towards higher energies. In addition, the shape of the upper polariton is slightly modified --- its onset reflects the convergent phase-space power law $\omega^3$ of Eq.~(\ref{Abspowerlaw}) found for the 2DEG absorption. This is emphasized in the inset. 
At zero cavity detuning [Fig.~\ref{combined_cuts}(b)], the situation of the finite and infinite mass cases is qualitatively similar.
When the cavity detuning is further increased, the position of the pole-like lower polariton approaches the direct threshold at $\omega = \beta\mu$ (indicated by the vertical dotted line). When the pole is in the energy interval $[0,\beta\mu]$, the lower polariton overlaps with the 
2DEG continuum absorption, and is therefore broadened. This is clearly seen in  Fig. \ref{combined_cuts}(c): Instead of a sharp feature, there is just a small remainder of the lower polariton at $\omega = \beta\mu$. 
As a result, one may say that in the regime of the Fermi-edge singularity, i.e., large $\mu$, the finite mass will cut off the excitonic features from the polariton spectrum  -- instead of the avoided crossing of Fig.~\ref{pureexcitonpolariton}, Fig.~\ref{combined_spectra}(b) exhibits an almost photonic-like spectrum, with a small (cavity) linewidth below the threshold at $\omega = \beta\mu$, and a larger linewidth above the threshold, reflecting the step-like 2DEG absorption spectrum of Fig.~\ref{FEScomp}.
The finite mass thus leads to a general decrease of the mode splitting between the two polariton branches. This trend continues 
when the Fermi energy is increased further. 

It is instructive to compare this behavior with the experimental results reported in~\cite{Smolka}. There, two differential reflectivity measurements were conducted, which can be qualitatively identified with the polariton spectra. The first measurement was carried out using a low-mobility GaAs sample (which should behave similarly to the limit of large VB hole mass), and moderate Fermi energies. A clear avoided crossing was seen, with the upper polariton having a much larger linewidth than the lower one (see Fig.~2(A) of \cite{Smolka}). In the second measurement, the Fermi energy was increased further, and a high-mobility sample was studied, corresponding to finite mass. A substantial reduction of the mode splitting between the polaritons was observed (Fig.~2(C) of \cite{Smolka}). While a detailed comparison to the experiment of \cite{Smolka} is
challenging, due to the approximations we made and the incongruence of the parameter regimes (in the experiment one has $\mu \simeq E_B$), the general trend of reduced mode splitting is correctly accounted for by our theory. 

\subsection{Small Fermi energy}
We now switch to the regime of of small Fermi energy discussed in Sec.~\ref{Photon self-energy small mu sec}, a regime in which the polariton spectra have not been studied analytically before. We again assume that the condition~ (\ref{M0smallerthanmu}), required for the approximating 
the photon self-energy by Eq.~(\ref{Kubo-formula}), is fulfilled. This may be appropriate for systems with a large exciton-binding energies,  e.g., transition metal dichalcogenide monolayers as recently studied in \cite{sidler2017fermi}.

\begin{figure}[H] 
\centering
\includegraphics[width=\columnwidth]{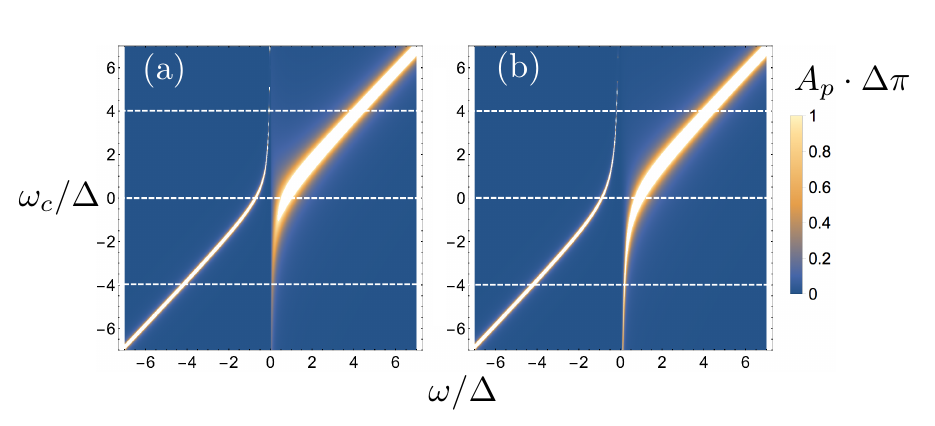}
\caption{(Color online) $\mu \ll E_B$: Exciton-polariton spectrum for small Fermi energy.
The white dashed lines denote the location spectral cuts presented in Fig.~\ref{combined_cuts2}.
(a) Infinite mass. This plot was obtained by inserting the Exciton Green's function for $\mu\gtrsim 0$, given by Eq.~(\ref{Nozieresresult}) multiplied by $\Delta^2 =  d_0^2\rho E_B/g^2$, into the photon Green's function, Eq.~(\ref{Polaritonspectralfunction}). Parameters: $\mu = 10 \Delta$, $\Gamma_c = 0.01\Delta$, $\alpha^2 = (\delta/\pi -1)^2 = 0.25$.
(b) Finite mass, with mass ratio $\beta=0.4$. In this plot, the finite mass Exciton Green's function, Eq.~(\ref{Piexcfinitemass}), was used, with the same parameters as in (a). 
} 
\label{combined_spectra2}
\end{figure}

For infinite mass, we may use Eq.~(\ref{Nozieresresult}) as photon self-energy, multiplied by a prefactor $\Delta^2 = d_0^2 \rho E_B/ g^2 $ [cf.\ Eq.~(\ref{Deltadef})], and expand the real and imaginary parts to leading order in $\alpha^2 = (\delta/\pi-1)^2$. The energy $\omega$ is now measured from the exciton pole: $\omega = \Omega-\Omega_T^{\text{exc}}$, $\Omega_T^{\rm{exc}} = E_G + \mu - E_B$. The corresponding polariton spectrum for a small cavity linewidth is shown in Fig.~\ref{combined_spectra2}(a). Qualitatively, it strongly resembles the bare exciton case as in Fig.~\ref{pureexcitonpolariton} (note that in Fig.\ \ref{combined_spectra2} the cavity linewidth was chosen to be 100 times smaller than in Fig.~\ref{pureexcitonpolariton}), but with a larger linewidth of the upper polariton. This is due to the possible polariton decay into the particle hole continuum contained in the excitonic power law, Eq.~(\ref{Nozieresresult}).

\begin{figure*}[!]
\centering
\includegraphics[width=\textwidth]{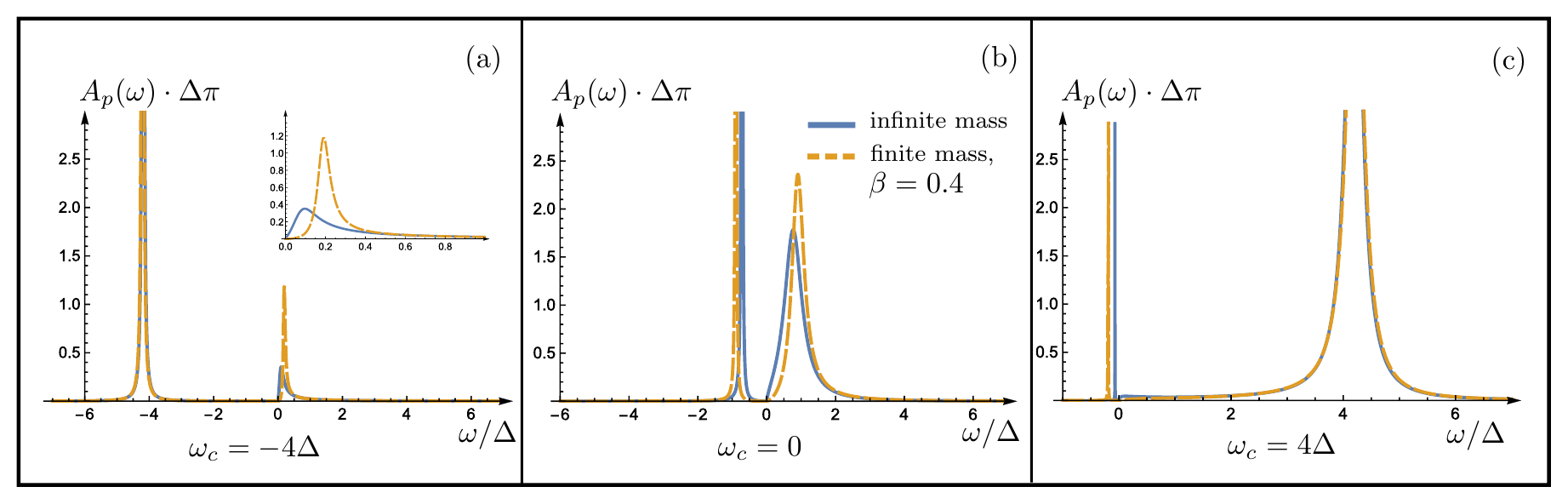}
\caption{(Color online) $\mu \ll E_B$: Spectral cuts at fixed cavity detuning through the polariton spectra of Fig.~\ref{combined_spectra2}, for both infinite (continuous blue lines) and finite hole mass (dashed orange lines). 
(a) Large negative cavity detuning. The inset shows a zoom onto the upper polaritons. 
(b) Zero cavity detuning. 
(c) Large positive cavity detuning. }
\label{combined_cuts2}
\end{figure*}

The detailed discussion of polariton properties in the regime of $\mu \ll E_B$ parallels the previous discussion in the regime $E_B \ll \mu$. For small negative detuning $\omega_c$ [Fig.\ \ref{combined_cuts2} (a)], the lower polariton is found at approximately $\omega = \omega_c$. The upper polariton has a significantly smaller weight, its shape reflects the excitonic power law of Eq.~(\ref{Nozieresresult}). However, compared to the previous spectral cuts (Fig.~\ref{combined_cuts}) the upper polariton peak is much more pronounced. This results from the exciton being now pole-like, 
as compared to the power law Fermi-edge singularity.  Increasing the detuning, weight is shifted to the upper polariton. At zero detuning [Fig.~\ref{combined_cuts2}(b)], the weight of the lower polariton is only order $\mathcal{O}\left(\alpha^2\right)$ larger than the weight of the upper polariton.  At large positive detuning, the position of the lower polariton is found at approximately 
\begin{align}
\label{peaklargewc_no}
\omega \sim -\omega_c^{-1/(1-\alpha^2)} \quad \text{as} \quad \omega_c \rightarrow \infty.
\end{align}
The lower polariton thus approaches the exciton line faster than in the pure exciton case, but slower than in the Fermi-edge regime [Eq.~(\ref{peaklargewc})].
A similar statement holds for the weight of the lower polariton, which scales as $\omega_c^{-2-\alpha^2}$.

The spectrum in the finite mass case is qualitatively similar, see Fig.~\ref{combined_spectra2}(b). 
Quantitatively, a stronger peak repulsion can be seen, which may be attributed to the enhanced excitonic quasiparticle weight in the finite mass case. 
A comparison of spectral cuts in the finite mass case [Fig.~\ref{combined_cuts2}(a)--(c)] further corroborates this statement [especially in 
Fig.~\ref{combined_cuts2}(c)]. Indeed, one finds that the position of the lower polariton at large cavity detuning is approximately given by 
\begin{align}
\label{peaklargewc_nofinite}
\omega \sim -\beta^{\alpha^2} \cdot \omega_c^{-1} \qquad 
\text{as} \quad \omega_c \rightarrow \infty , 
\end{align}
i.e., the excitonic line at $\omega =0$ is approached more slowly than in the infinite mass case, Eq.~(\ref{peaklargewc_no}). The corresponding weight  falls off as $\omega_c^{-2}$. Thus, the lower polariton has a slightly enhanced weight compared to the infinite mass case. 
In addition, in the spectral cut at large negative detuning, [inset to Fig.~\ref{combined_cuts2}(a)], the upper polariton appears as a sharper peak compared to the infinite mass case, which again results from the enhanced quasi particle weight of the finite mass case.

\section{Conclusion}
\label{Conclusion sec}
In this paper we have studied the exciton-polariton spectra of a 2DEG in an optical cavity in the presence of finite CB electron density.
In particular, we have elucidated the effects of finite VB hole mass, distinguishing between two regimes.
In the first regime (small Fermi energy as compared to the exciton binding energy), we have found that excitonic features in the 2DEG absorption are enhanced by the exciton recoil and the resulting suppression of the Fermi edge singularity physics. In contrast, in the second regime of Fermi energy larger than the exciton binding energy, 
it is the VB hole which recoils at finite mass. This cuts off the excitonic features. These modifications also translate to polariton spectra, especially to the lower polariton at large cavity detuning, which is exciton-like. Our findings reproduce a trend seen in a recent experiment~\cite{Smolka}.

We would like to mention several possible extensions of this work.
To begin with, it would be promising to study the effect of long-range interactions on the power laws, and hence on polariton spectra, from an analytical perspective. Long-range interactions are expected to be most important in the regime of small Fermi energy, leading to additional bound states and to the Sommerfeld enhancement effects~\cite{haug1990quantum}. Moreover, one should try to explore trionic features, for which it is necessary to incorporate the spin degree of freedom (to allow an electron to bind to an exciton despite the Pauli principle).
Another interesting direction would be to tackle the limit of equal electron and hole masses, which is relevant to transition metal dichalcogenides, whose polariton spectra in the presence of a Fermi sea where measured in a recent experiment~\cite{sidler2017fermi}. Lastly, one should address the behavior of the polariton in the regime of small Fermi energy and strong light-matter interactions. Then, not the exciton, but rather the polariton interacts with the Fermi sea, and different classes of diagrams have to be resummed to account for this change in physics.

\begin{acknowledgments}
This work was initiated by discussions with A.~Imamo\u{g}lu.
The authors also acknowledge many helpful comments from F.~Kugler, A.~Rosch,  D.~Schimmel, and O.~Yevtushenko.
This work was supported by funding from the German Israeli Foundation (GIF) through I-1259-303.10.
D.P.\ was also supported by the German Excellence Initiative via the Nanosystems Initiative Munich (NIM). 
M.G.\ received funding from the Israel Science
Foundation (Grant 227/15), the US-Israel Binational
Science Foundation (Grant 2014262), and the Israel Ministry of Science and Technology (Contract 3-12419), while L.G.\ was supported by NSF Grant DMR-1603243.
\end{acknowledgments}

\appendix

\section{Evolution of absorption spectra with increasing chemical potential}
\label{muincapp}

In this Appendix, we present an extended overview of how the absorption spectra evolve inbetween the controlled extremal limits of $\mu \ll E_B$ and $\mu \gg E_B$.

For $\mu \ll E_B$, the dominant spectral feature is the exciton. For finite mass ($
\beta \neq 0$), it has a coherent delta-like part and an incoherent tail, see Eq.\ (\ref{bothfives}), while the infinite mass exciton ($\beta = 0$) is a purely incoherent power law, see Eq.\ (\ref{Aexcsummary}).  
These pronounced excitonic features are well separated from the CB continuum part at $\Omega_T^{\text{FES}} = E_G + \mu$ (see inset to Fig.\ \ref{finmasssmallmu1}).

As $\mu$ is increased, the incoherent exciton part [Eqs.\ (\ref{Exccases}) and (\ref{Aexcsummary})] starts to overlap with the CB continuum part. Moreover, the overall relative weight of both the coherent and incoherent portions of the exciton part of the spectrum (which are both proportional to $E_B$) will diminish. Still, within our simplified model which neglects CB electron-CB electron interactions, and for $\beta=0$, this exciton feature will never disappear completely, since in this model an infinite mass VB hole is simply a local attractive potential for the CB electrons, and such a potential will always have a bound state in 2D. However, for finite VB hole mass, the exciton energy (location of the coherent delta peak) will penetrate into the CB continuum when $\mu$ becomes larger than $E_B/\beta \gg E_B$ (i.e., when $E_B$ crosses the indirect threshold, see Fig.\ \ref{twothresholds}(a)). More importantly, CB electron-CB electron interactions would screen the hole potential, and will thus reduce the exciton binding energy and presumably eliminate the exciton part of the spectrum completely as soon as $\mu \gg E_B$.

\begin{figure}[H]
\centering
\includegraphics[width=\columnwidth]{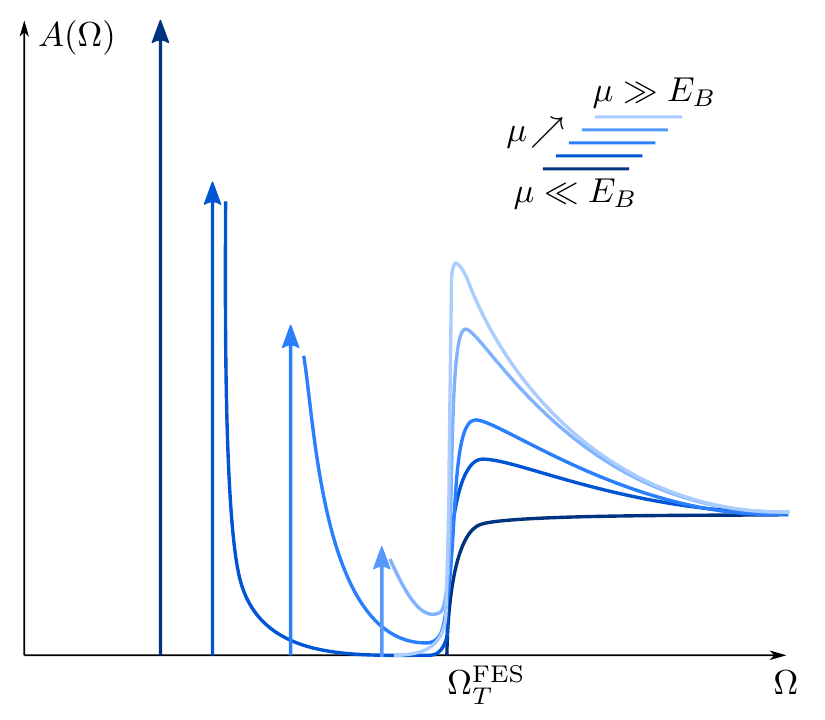}
\caption{(Color online): Putative evolution of absorption spectra as $\mu$ is increased. The colored arrows represent  delta-function peaks, their height corresponds to the relative weight of those peaks. The (hand-sketched) plots of this figure comprise the effects of a  (large) finite VB hole mass ($\beta \neq 0$) and electron-electron interactions, beyond what's actually computed in this paper. 
For clarity, the shift of the spectra with increasing $\mu$ is disregarded. 
For $\mu$ even larger than shown in the sketch, the FES will reduce to a step-like feature again.}
\label{increased_mu}
\end{figure}

 To describe this situation, it has been customary in the literature \cite{combescot1971infrared,PhysRevB.35.7551} to still employ the same simplified model neglecting CB electron-CB electron interactions, but assume that the hole potential does not create a bound state for large enough $\mu$, a practice we follow in this work as well. Then, for $\mu \gg E_B$, one should concentrate on the remaining, CB continuum part of the spectrum, which will evolve into the Fermi-edge singularity (FES), cut off by the VB hole recoil energy for $\beta \neq 0$. 
A putative evolution of absorption spectra with increasing $\mu$ is sketched in Fig.\ \ref{increased_mu}.

\section{Evaluation of the exciton self-energy in the time-domain}
\label{technical}

In this Appendix, we present the time-domain evaluation of the exciton self-energy diagrams of Fig.~\ref{directtimedomain}. These diagrams contain one CB electron loop only, and therefore yield the leading contribution when $\mu/E_B$ is small.
We will start with the direct diagrams [Fig.~\ref{directtimedomain}(a)], and then turn to the exchange series [Fig.~\ref{directtimedomain}(b)].

\subsection{Direct diagrams}
\begin{figure}[H]
\centering
\includegraphics[width=\columnwidth]{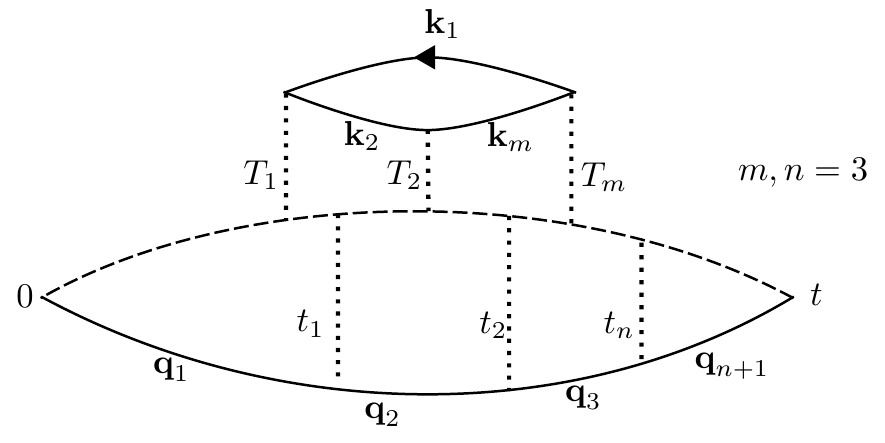}
\caption{A direct self-energy diagram in the time-domain. The Green's function with an arrow indicates the CB electron propagating backwards in time. }
\label{directtimedomainapp}
\end{figure}

First, we note that the bare Green's functions in the time domain read
\begin{align} 
G^{(0)}_c(\textbf{k},t) &= -i(\theta(t) - n_\textbf{k})e^{-i\epsilon_\textbf{k}t}, \\
G^{(0)}_v(t) &= i\theta(-t) e^{iE_Gt},
\end{align} 
with the zero temperature Fermi function $n_\k = \theta(k_F-k)$. 
Using these, we will evaluate the series of direct diagrams of Fig.~\ref{directtimedomain}(a). The temporal structure of a generic direct diagram is illustrated via the example of Fig.~\ref{directtimedomainapp}. 
\vspace{1em}

To compute such a diagram, we make the following observation: Since the VB propagator has no momentum dependence, all VB phase factors simply add up to give a total factor of $ e^{-iE_Gt} $. Then, the step functions in the VB propagators enforce time ordering for the intermediate time integrals. In the specific case shown in Fig.~\ref{directtimedomainapp}, $0<T_1<t_1<T_2<t_2<T_m<t_n<t$ with $m = n =3$ ($m$ and $n$ count the number of interaction lines above and below the dashed VB line, respectively).
However, there are also diagrams with $m=n=3$, but with a different relative ordering of the interaction lines.  Summing over all those diagrams for $m$ and $n$ fixed, one needs to integrate over the time ranges $0<t_1<...<t_n<t \cap 0<T_1<...<T_m<t$.
This means that the time integration for the direct diagrams splits into a product of two functions, representing the propagation of a Fermi sea electron (above the VB line in Fig. \ref{directtimedomainapp}) and a photoexcited electron (below the VB line) in the time-dependent potential.

We are now in the position to write down the full expression for the sum of direct diagrams $D$ to all orders in the interaction, fixing the signs with Wick's theorem:
\begin{align}
\label{longappd}
& D(t) = -\int_{k_1<k_F} \frac{d\textbf{k}_1}{(2 \pi)^2} \ e^{-iE_Gt} \tilde{B}(t) C(t),
\end{align}
where
\begin{widetext}
\begin{align}
\label{longapp}
& \tilde{B}(t) =
 \sum_{m =1}^\infty (-V_0)^m
 \hspace{-2em}\int\displaylimits_{0<T_1<\cdots<T_m<t} \hspace{-2.5em} dT_1 \cdots dT_m
 \hspace{-.8em} \int\displaylimits_{k_2>k_F}  \hspace{-1em}  \frac{d\textbf{k}_2}{(2 \pi)^2} \cdots \hspace{-.8em} \int\displaylimits_{k_m>k_F} \hspace{-1em} \frac{d\textbf{k}_m}{(2 \pi)^2}  
 \tilde{G}_c(\textbf{k}_1, T_1-T_m) \tilde{G}_c(\textbf{k}_2, T_2-T_1) \cdots \tilde{G}_c(\textbf{k}_m, T_m - T_{m-1}),
 \\ 
& C(t) = \sum_{n=0}^{\infty} (-V_0)^n
\hspace{-2em}\int\displaylimits_{0<t_1<\cdots<t_n<t}  \hspace{-2.5em}dt_1 \cdots dt_n
\hspace{-.8em}\int\displaylimits_{q_1>k_F} \hspace{-1em} \frac{d\textbf{q}_1}{(2 \pi)^2} \cdots \hspace{-.8em} \hspace{-.8em}\int\displaylimits_{q_{n+1}>k_F}  \hspace{-1em} \frac{d\textbf{q}_{n+1}}{(2 \pi)^2}
\tilde{G}_c(\textbf{q}_1,t_1) \tilde{G}_c(\textbf{q}_2,t_2-t_1) \cdots \tilde{G}_c(\textbf{q}_{n+1}, t-t_n),
\end{align}
\end{widetext}
and
\begin{align}
\tilde{G}_c(\textbf{k}_1, T_1-T_M) &= ie^{-i\epsilon_{\k_1}(T_1-T_M)} \\ \nonumber 
\tilde{G}_c(\textbf{p},\tau) &= -ie^{-i\epsilon_\textbf{p}\tau} \qquad \text{for} \quad \textbf{p} \neq \textbf{k}_1 \ .
\end{align}
Defining the retarded Green's function by
\begin{align} 
G^{0, R}_c(\textbf{p},\tau) = \theta(\tau) \tilde{G}(\textbf{p},\tau),
\end{align}
we can rewrite the two factors appearing in $D(t)$ as sequences of convolutions:
\begin{align} & 
\label{Aconvolution}
B(t) \equiv \\ \notag
&e^{-i\epsilon_{\k_1}t} \tilde{B}(t) =  \sum_{m=1}^{\infty} (-V_0)^m \hspace{-.8em}\int\displaylimits_{k_2>k_F} \hspace{-1em} \frac{d\textbf{k}_2}{(2 \pi)^2} \cdots \hspace{-.8em}\int\displaylimits_{k_{m}>k_F}\hspace{-1em} \frac{d\textbf{k}_{m}}{(2 \pi)^2} \\ &\left[G_c^{0,R}(\textbf{k}_1,\ \!) \ast G_c^{0,R}(\textbf{k}_2,\ \!) \cdots  \ast G_c^{0,R}(\textbf{k}_{m},\ \!)\ast G_c^{0,R}(\textbf{k}_1,\ \!)\right]\!(t),\notag
\\ & \label{Bconvolution}
C(t) = \sum_{n=0}^{\infty} (-V_0)^n \hspace{-.8em}\int\displaylimits_{q_1>k_F} \hspace{-1em} \frac{d\textbf{q}_1}{(2 \pi)^2} \cdots \hspace{-.8em}\int\displaylimits_{q_{n+1}>k_F} \hspace{-1em}\frac{d\textbf{q}_{n+1}}{(2 \pi)^2} \\ \notag &\left[G_c^{0,R}(\textbf{q}_1,\ ) \ast \cdots \ast  G_c^{0,R}(\textbf{q}_{n+1},\ )\right](t).
\end{align}
Together, Eqs.~(\ref{longappd}) and (\ref{Aconvolution})--(\ref{Bconvolution}) correspond to Eq.~(\ref{Dnew}) in the main text.
Fourier transforming Eq.~(\ref{longappd}) results in: 
\begin{align} 
\label{Dconvo}
D(\Omega) = \hspace{-.8em}\int\displaylimits_{k_1<k_F}\hspace{-.8em} \frac{d\textbf{k}_1}{(2 \pi)^2} \underbrace{ i \int \frac{d\nu}{2\pi} B(\nu) C(\Omega-E_G + \epsilon_{\k_1} - \nu)}_{I(\Omega)},
\end{align}
where we defined $I(\Omega)$ for later purpose. The Fourier transform of $B(t)$ reads:
\begin{align} 
&
\label{Anu_app}
B(\nu) = \sum_{m=1}^{\infty} (-V_0)^m \int\displaylimits_{k_2>k_F} \frac{d\textbf{k}_2}{(2 \pi)^2} \cdots \int\displaylimits_{k_{m}>k_F} \frac{d\textbf{k}_{m}}{(2 \pi)^2} \\ &  G_c^{0,R}(\textbf{k}_1,\nu ) \cdot G_c^{0,R}(\textbf{k}_2,\nu ) \cdots G_c^{0,R}(\textbf{k}_{m},\nu )\cdot G_c^{0,R}(\textbf{k}_1,\nu ),
\notag
\end{align} 
with retarded real frequency Green's functions:
\begin{align} 
\label{Grapp}
G_c^{0,R}(\textbf{k},\nu) = \frac{1}{\nu - \epsilon_{\textbf{k}} + i0^+}.
\end{align}
Inserting (\ref{Grapp}) into (\ref{Anu_app}), the integrations are trivially performed. The summation over interaction lines reduces to a geometric series, yielding: 
\begin{align}
\label{Alog_app}
B(\nu) &= \frac{-V_0}{g}  \frac{1}{(\nu - \epsilon_{\k_1} + i0^+)^2}\cdot \frac{1}{\ln\left(\frac{\nu - \mu + i0^+}{-E_B}\right)},
\end{align}
where we used $\ln(E_B/\xi) = -1/g$, c.f.\ Eq.~(\ref{EBfctofg}).
For the term $C(\Omega - E_G + \epsilon_{\k_1} - \nu)$ appearing in (\ref{Dconvo}) we analogously arrive at: 
\begin{align}
\notag
&C(\Omega - E_G + \epsilon_{\k_1} - \nu) =  \frac{\rho}{g}\left(1- \frac{1}{g \ln\left(\frac{\kappa- \nu + i0^+}{-E_B}\right)}\right), \\  &\qquad  \kappa \equiv \Omega - E_G  + \epsilon_{\k_1} - \mu.
\end{align}

The functions $B(\nu)$ and $C(\nu)$ are difficult to integrate, because 
they each have both a pole and a branch cut, arising from the $1/\ln$ term. We can split these terms as follows:
\begin{align}
\label{polebranchsplit} 
& \frac{1}{\ln\left(\frac{\nu - \mu + i0^+}{-E_B}\right)} =  \notag  \frac{-E_B}{E_B + \nu - \mu + i0^+} \\ &+ \left(\frac{1}{\ln\left(\frac{\nu - \mu + i0^+}{-E_B}\right)} + \frac{E_B}{E_B + \nu - \mu + i0^+}\right).
\end{align}  
The first term on the right hand side of Eq.~(\ref{polebranchsplit}) has just a simple pole, while the second one's only singularity is a branch cut. Using this representation, we can  evaluate $I(\Omega)$ as defined in Eq.~(\ref{Dconvo}) employing the following argument: Physically, the terms $B$, $C$ represent the propagation of the two electrons in the hole potential. Comparing to the simple exciton ladder summation (see Sec.~\ref{Photon self-energy zero mu sec}), 
we associate the poles of the $1/\ln$-terms in these functions with the exciton contribution, while the branch cut corresponds to the continuum above the indirect threshold, $\Omega > E_G + \mu$. 

Following these observations, let us split $I(\Omega)$ into a pole-pole, a pole-branch, and a branch-branch contribution.
$I_{\text{branch-branch}}$ only contributes to the continuum part of the spectrum. More importantly (as explained in the main text), employing spectral representations of the retarded functions $B_\text{branch}, C_\text{branch}$, it is easily shown that $\text{Im}\left[I_{\text{branch-branch}}\right]$
(which is of potential importance for the lineshape of the exciton spectrum)  vanishes for frequencies close to the exciton pole ($\omega \gtrsim 0$).  It is thus not important for our purposes.

Computing contour integrals, $I_{\text{pole-pole}}$ is easily evaluated to give:
\begin{align}
\label{trionpole} 
I_{\text{pole-pole}}(\omega) = \frac{E_B^2}{g^2}\frac{1}{(\omega + i0^+)^2}\frac{1}{E_B + \omega + \epsilon_{\textbf{k}_1}- \mu + i0^+},
\end{align}
where energies are measured from the exciton pole, $\omega = \Omega - (E_G + \mu) + E_B$. This contribution gives rise to trionic features in the spectrum, which are shortly discussed in Appendix~\ref{trion-contribution}.

Last, computing contour integrals and disregarding terms which are subleading in $\omega/E_B$, the pole-branch contribution is found to be:
\begin{align} 
\label{Ipolebranch} \notag
&I_{\text{pole-branch}}(\omega) \simeq \frac{-E_B}{g^2}\frac{1}{(\omega + i 0^+)^2}\times \\ &\left(\frac{1}{\ln\left(\frac{\omega + \epsilon_{\k_1} - \mu + i0^+}{-E_B}\right)} + \frac{E_B}{E_B + \omega + \epsilon_{\k_1} - \mu + i0^+}\right).
\end{align} 
Inserting  the Eqs.~(\ref{trionpole}) and~(\ref{Ipolebranch}) into Eq.~(\ref{Dconvo}), one finally arrives at Eq.~(\ref{Ddirectfinal}) of the main text.

\subsection{Exchange diagrams}
\begin{figure}[H]
\centering
\includegraphics[width=\columnwidth]{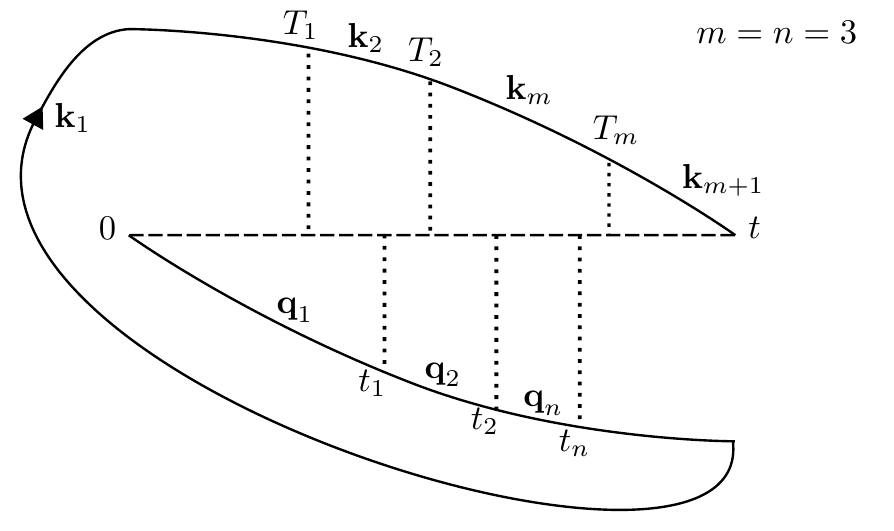}
\caption{An exchange self-energy diagram in the time-domain. The Green's function shown with an arrow indicates the CB electron propagating backwards in time.}
\label{exchangetimedomainapp}
\end{figure}

The computation of the exchange diagrams, though technically sligthly more involved, essentially proceeds along the same lines. The general time-structure of an exchange diagram is illustrated in Fig.~\ref{exchangetimedomainapp}.
\vspace{0.2cm}

As for the direct diagrams, the VB propagators just enforce a time ordering. In addition, there is the condition
$t_n>T_1$.
When this condition is violated, the diagram reduces to a ladder diagram, which must be excluded to avoid double counting. 
Taking this into account, the full expression for the sum of exchange diagrams reads:
\begin{widetext} 
\begin{align}
X(t) =  \notag
\sum_{m,n=1}^{\infty} \!(-V_0)^{m+n} & e^{-iE_G t}
\hspace{-2em}\int\limits_{0<T_1<\cdots<T_m<t} \hspace{-2.3em}dT_1 \cdots dT_m \int\limits_{T_1}^t dt_n \int\limits_0^{t_n} dt_{n-1} \cdots \int\limits_0^{t_2} dt_1
 \hspace{-.8em} \int\limits_{k_1<k_F} \hspace{-.7em}\frac{d\textbf{k}_1}{(2\pi)^2}\hspace{-.7em}\int\limits_{k_2>k_F} \hspace{-.7em}\frac{d\textbf{k}_2}{(2 \pi)^2} \cdots \hspace{-1.2em}\int\limits_{k_{m+1}>k_F}\hspace{-1em} \frac{d\textbf{k}_{m+1}}{(2 \pi)^2} \hspace{-.7em}   \int\limits_{q_1>k_F} \hspace{-1em} \frac{d\textbf{q}_1}{(2 \pi)^2} \cdots \hspace{-.7em} \int\limits_{q_{n}>k_F} \hspace{-1em}\frac{d\textbf{q}_{n}}{(2 \pi)^2}
\\   & \tilde{G}_c(\textbf{k}_1, T_1 - t_n) \tilde{G}_c(\textbf{k}_2, T_2 - T_1) \cdots \tilde{G}_c(\textbf{k}_{m+1}, t - T_m) \tilde{G}_c(\textbf{q}_1, t_1)\cdots \tilde{G}_c(\textbf{q}_n,t_n-t_{n-1}) 
\label{Xtapp}
\end{align}
 
To rewrite (\ref{Xtapp}) as a sum of convolutions, one can employ the following easily-derived formula:
\begin{align}
\label{genconv}
&\mathcal{F}\left(\int_{-\infty}^{\infty} dt_1 f(t-t_1)  g(t,t_1)\right)(\Omega) = \int_{-\infty}^{\infty} \frac{d\omega_1}{2\pi} f(\omega_1) g(\Omega - \omega_1, \omega_1),
\end{align}
\end{widetext}
where $\mathcal{F}$ denotes the Fourier transform, and $f$ and $g$ are any two well-behaved functions. 
Applying this result, a computation similar to the one for $D(\Omega)$ shows that the Fourier-transform of Eq.~(\ref{Xtapp}) can be expressed as: 
\begin{widetext}
\begin{align}
\nonumber
&X(\Omega) = -\int_{k_1< k_F} \frac{d\textbf{k}_1}{(2\pi)^2} \int_{-\infty}^{\infty} \frac{d\omega_1}{2\pi} (-g) \ln\left(\frac{\omega_1 - \mu + i0^+}{-\xi}\right) \frac{1}{1+g \ln\left(\frac{\omega_1 - \mu + i0^+}{-\xi}\right)} \frac{1}{\Omega - E_G - \omega_1 + i0^+} \\ & \int_{-\infty}^{\infty}\frac{d\omega_2}{2\pi}(-g) \ln\left(\frac{\omega_2 - \mu + i0^+}{-\xi}\right)\frac{1}{1+g \ln\left(\frac{\omega_2 - \mu + i0^+}{-\xi}\right)} \frac{1}{-\omega_2 + \Omega - E_G + i0^+} \frac{1}{\omega_2 + \omega_1 - \Omega + E_G - \epsilon_{\textbf{k}_1} - i0^+}.
\label{Xsomewhere}
\end{align} 
\end{widetext}
This expression can be evaluated as before, splitting it into pole-pole, pole-branch and branch-branch contributions using Eq.~(\ref{polebranchsplit}). In complete analogy to the direct diagrams, the imaginary part of the branch-branch contribution can be shown not to contribute in the regime of interest to us, and we therefore disregard it completely. Straight-forwardly evaluating the pole-pole and pole-branch contributions, one ultimately arrives at Eq.~(\ref{Xomegamaintext}) in the main text.

\section{Trion contribution to the exciton self-energy diagrams} 
\label{trion-contribution}

The pole-pole contribution to the direct self-energy $D(\omega)$ [Eq.~(\ref{Dnew})] physically represents two electrons tightly bound to the hole potential. Indeed, it assumes the form:
\begin{align}
\label{DandI} 
D_{\text{pole-pole}}(\omega) = \int_{k_1<k_F} \frac{d\textbf{k}_1}{(2 \pi)^2} I_{\text{pole-pole}}(\omega),
\end{align}
where $I_{\text{pole-pole}}$ is given in Eq.~(\ref{trionpole}). $I_{\text{pole-pole}}$ can be identified with a bare trion Green's function, since it has a pole at $\omega = -E_B + \mu-\epsilon_{\textbf{k}_1}$, corresponding to the binding of a second CB electron to the exciton (recall that $\omega$ is measured from the exciton threshold), where the energy $\epsilon_{\textbf{k}_1}$ of this second electron 
can be from anywhere in the Fermi sea. Evaluation of (\ref{DandI}) close to the trion resonance $\omega \simeq -E_B$ leads to 
\begin{align} 
\label{Dappevaluated}
D_{\text{pole-pole}}(\omega) \simeq  \frac{\rho}{g^2} \ln\left(\frac{E_B + \omega + i0^+}{E_B + \omega - \mu}\right).
\end{align}
Using Eq.\ (\ref{SigmaDint}) of the main text, (\ref{Dappevaluated}) gives rise to a self-energy contribution to the exciton
\begin{align}
\label{trionselfapp}
\Sigma_{\text{exc}}= E_B \ln\left(\frac{E_B + \omega + i0^+}{E_B + \omega - \mu}\right).
\end{align}
This self-energy expression fully matches usual results found in works concerned with trions \cite{suris2001excitons, sidler2017fermi, efimkin2017many}, apart from two minor differences: First, in these works the case of finite VB hole mass (of the same order as the CB mass) is considered, but reevaluation of (\ref{trionselfapp}) for finite mass is straightforward and only results in some trivial factors involving mass ratios. Second, in the works cited above the exciton is treated as an elementary entity, and the trion binding energy is therefore an adjustable parameter. By contrast, we have started from a microscopic model which does not contain excitons, and, accounting for exchange processes, computed excitons and trions along the way. As a result, our microscopic theory yields the same binding energy $E_B$ for excitons and trions. However, this is clearly an artefact of disregarding electron-electron interactions (which would significantly reduce the trion binding energy), and can heuristically be accounted for by replacing $E_B$ in Eq.\ (\ref{trionselfapp}) by a trion binding energy $E_T \ll E_B$. 
Upon inserting (\ref{trionselfapp}) into the exciton Green's function (\ref{dressedexcwithsigma}), one finds the following spectral features: First, there is a sharp resonance, red detuned w.r.t.\ the trion threshold by an order of $\mu$, and with a weight that scales as $\mu/E_T$. This peak is commonly called the trion, or, more appropriately, attractive polaron \cite{sidler2017fermi}, since the trion bound state is not filled. Second, there is a small step-like feature for $0< E_B + \omega < \mu$, arising from the imaginary part of (\ref{trionselfapp}). This feature, where the trion bound state is filled and the second electron constituting the trion can come from anywhere in the Fermi sea, has smaller (but not parametrically smaller) weight than the attractive polaron, and is usually overlooked in the literature. Investigation of further trion properties is a worthwhile goal which we leave for further work. 

Let us close this Appendix with a technical remark: Of course, for spinless electrons a trion cannot exist in our simple model of short range VB hole-CB electron interaction, due to the Pauli principle (two electrons cannot occupy the single bound state created by the hole). In line with that, the pole-pole contribution cancels in this case between the direct and exchange diagrams. However, in the spinful case, the direct contribution will incur a factor of two, so it does not cancel with the exchange contribution, so the trion remains.

\section{The self-energy contribution of the exchange diagrams}
\label{Fumidiscussion}

The exchange contribution to the exciton self-energy, Eq.~(\ref{Fumitypeshift}), can be understood by the following considerations.
The ground state energy of an $N$-particle system in the presence of an attractive delta function potential strong enough to form a bound state is lower than the $N$-particle ground state energy of the system without the potential by an amount
\begin{align}
\Delta E = -E_B - (1-\alpha)\mu,
\end{align}
which is the sum of the bound state energy $E_B$, and a second term which arises from the rearrangement of the Fermi sea, described by Fumi's theorem~\cite{G.D.Mah2000} [recalling that $1-\alpha=\delta/\pi$, cf.\ Eq.~(\ref{alphaisphase})].
We find that the exchange diagrams give the contribution $\mu$, while the term $\alpha\mu$ stems from the direct diagrams [Eq.~(\ref{sigmadendlich})]. 
To create such an attractive potential, one has to lift one electron from the VB to the CB, which costs $E_G + \mu$. In our treatment, the extra cost $\mu$ appearing here is contained in the shift of the pole of the ladder diagrams, Eq.~(\ref{muexcitonpole}). Thus, the minimal absorption energy predicted by our model is $E_G - E_B + \alpha \mu \approx E_G - E_B$.

At first sight this seems to contradict the experimental results (e.g., \cite{sidler2017fermi}), according to which the minimal absorption energy is $E_G - E_B + \mu$ (or $2\mu$ for equal electron-hole masses). This is attributed to ``phase-space filling effects'', or, in other words, the Burstein-Moss shift \cite{PhysRev.93.632}, which precisely correspond to the shift of the ladder pole, without the Fumi contribution. 
The reason for this discrepancy is that our model ignores the CB electron-CB electron interaction, which would render the exciton electrically neutral and suppress the Fumi shift. Thus, as also pointed out in the literature on the X-ray edge problem, neglecting electron-electron interactions gives the right power law scalings of the spectra only, but not the correct threshold energies. 

Another aspect of Eq.~(\ref{Fumitypeshift}) is its lack of dependence on the frequency $\omega$. In other words, the Anderson orthogonality power law of the exciton Green's function does not depend on $X(\omega)$.
This could have been anticipated 
by an argument based on Hopfield's rule of thumb \cite{Hopfield1969} and the  
results of \cite{combescot1971infrared}.
Consider the spinful case, and study the absorption spectral function for, e.g., right-hand circularly polarized light at the exciton threshold, creating a spin down electron and a spin up hole. The spectrum should have the form
\begin{align}
\label{shaky}
\frac{1}{\omega}\cdot \omega^{  (1-\delta_\downarrow/\pi)^2+(1-\delta_\uparrow/\pi)^2}.
\end{align}
For the spin down electrons, the exponent is $(1-\delta_\downarrow/\pi)^2$ rather than $(\delta_\downarrow/\pi)^2$ because of the Hopfield rule: one electron is lifted from the valence band to the conduction band. For the spin up electron, no electron is lifted. However, the exciton is the secondary threshold in the spinful case (the primary one is the trion). As seen from~\cite{combescot1971infrared}, the spin up exponent should therefore also be as in Eq.~(\ref{shaky}).
Now, in the spinful case all direct diagrams will come with a spin factor of 2, while the exchange diagrams will not. However,
we see that the exponent in (\ref{shaky}) is exactly 2 times the exponent the spinless case, Eq.~(\ref{Nozieresresult}), when recalling that
$\delta_\uparrow = \delta_\downarrow = \delta$ for our spin-independent potential. 
This shows that the exchange diagrams should indeed not contribute to Anderson orthogonality, at least to leading order. 

\section{Computation of phase-space integrals for the particle-hole pair density of states}
\label{phasespacesec}

To clarify the different role of the recoil in the exciton (section \ref{excfinitemasssec}) and FES cases (section \ref{FESfiniteholemasssubseq}), let us present the computation of two important phase space integrals. 

\subsection{Exciton recoil}
\label{excrecsec}

We start with the evaluation of the imaginary part of the exciton self-energy Im[$\Sigma$]($\omega$) given in Eq.\  (\ref{imende}), focusing on zero exciton momentum. Im[$\Sigma$] reads: 
\begin{align} 
\label{phasespacefirst}
\nonumber
&\text{Im}[\Sigma_{\text{exc}}] \simeq  -\frac{\pi V_0}{\rho g} \alpha^2 \int_{k_1 < k_F} \frac{d\textbf{k}_1}{(2\pi)^2}  \int_{k_2 > k_F} \frac{d\textbf{k}_2}{(2\pi)^2}  \\   &\qquad\qquad \delta(\omega - (\k_2 - \k_1 )^2/2M_\text{exc} - \epsilon_{\textbf{k}_2} + \epsilon_{\textbf{k}_1}) .
\end{align}

 Im[$\Sigma_{\text{exc}}$] can be interpreted as rate of decay of excitons into CB electron-hole pairs, or alternatively as density of state of the CB pairs.
We aim to compute the leading $\omega$-behaviour of Im[$\Sigma_{\text{exc}}$]. To put it short, the delta-function in (\ref{phasespacefirst}) requires $\k_1, \k_2 \simeq k_F$ and $\measuredangle(\k_1, \k_2) \simeq 0$, and these phase space restrictions pile up to give Im[$\Sigma_{\text{exc}}] \sim \omega^{3/2}$. To perform the calculation in detail, we substitute $
\textbf{x} = \frac{\textbf{k}_2}{\sqrt{2m}} , \ \textbf{y} = \frac{\textbf{k}_1}{\sqrt{2m}} 
$.
Switching the integrals for convenience, we can rewrite (\ref{phasespacefirst}), to leading order in the mass ratio $\beta$, as
\begin{align} 
\label{firstsubst}
\text{Im}[\Sigma_{\text{exc}}] = -&\frac{\alpha^2}{\pi} \int_{x>\sqrt{\mu}}  d\textbf{x} \int_{y< \sqrt{\mu}} d\textbf{y}\\  \notag  &\delta\left(\omega -  (x^2 - \mu) + (y^2 - \mu)  - \beta(\textbf{x}- \textbf{y})^2 \right)  .
\end{align} 
First, it is obvious that (\ref{firstsubst}) is proportional to $\theta(\omega)$, since all terms subtracted from $\omega$ in the delta function are positive, hence there cannot be any cancellations.
Second, it is clearly seen that $x\simeq \sqrt{\mu}$, $y \simeq \sqrt{\mu}$ to yield a nonzero contribution for small $\omega$. Thus, we may linearize the dispersion relation, starting with $\textbf{y}$:
\begin{align} 
\label{linearize} 
\textbf{y} = (\sqrt{\mu} + \gamma_y)\textbf{e}_y, \\
y^2 = \mu + 2\sqrt{\mu}\gamma_y + \mathcal{O}(\gamma_y^2)  .
\end{align} 
In doing so, we effectively disregard subleading terms of order $\mathcal{O}(\omega^2/\mu)$ in the argument of the delta function. \\
Introducing the notation
\begin{align}
\phi=\measuredangle({\textbf{x}, \textbf{y}})  , \qquad c = \cos(\phi)  , 
\end{align} 
we arrive at: 
\begin{widetext}
\begin{align}  
&\text{Im}[\Sigma_{\text{exc}}] =\\ &-\frac{\alpha^2\theta(\omega)}{\pi}\hspace{-0.8em}\int\displaylimits_{x > \sqrt{\mu}} \hspace{-0.8em} d\textbf{x}  \int_{-1}^{1} \frac{2}{\sqrt{1-c^2}}\hspace{-0.4em} \int\displaylimits_{-\sqrt{\mu}}^0 \hspace{-0.6em}  d\gamma_y\ (\sqrt{\mu} + \gamma_y) \delta\bigg(\underbrace{\omega - (x^2-\mu) - \beta x^2 + 2 \beta x \sqrt{\mu} c - \beta \mu}_{=A} + \gamma_y \underbrace{\left(2\beta x c - 2\beta\sqrt{\mu} + 2\sqrt\mu\right)}_{=B} \bigg) \notag  .
\end{align} 
\end{widetext}
Since the only contribution comes from $\gamma_y$ close to the upper boundary, we can write $ \sqrt{\mu} + \gamma_y \simeq \sqrt{\mu} $. 
Using $B \simeq 2\sqrt{\mu}$, the trivial integral over $\gamma_y$ then results in 
\begin{align}
\label{dappsomewhere}
 \text{Im}[\Sigma_{\text{exc}}] =  -\frac{\alpha^2}{\pi} \int_{x > \sqrt{\mu}} d\textbf{x} \int_{-1}^1  dc \ \frac{1}{\sqrt{1-c^2}} \  \theta(A)  \ .
\end{align} 
To find the leading power law in $\omega$ of this expression, we assume that 
$\omega \ll \beta\mu $.
Then, we rewrite $\theta(A)$ as
\begin{align} 
\nonumber
&\theta(\overbrace{\omega - (x^2-\mu) - \beta x^2 - \beta \mu}^{=C} + 2 \beta x \sqrt{\mu} c ) =\\
 &\theta\left(c - (-C/2\beta x \sqrt{\mu} )\right)  .
\end{align} 
We now use $x \simeq \sqrt{\mu}$. Thus, we can write
\begin{align}
-C/2\beta x\sqrt\mu \simeq 1-\left(\frac{\omega}{2\beta\mu} - \frac{x^2 -\mu}{2\beta\mu}\right) + \mathcal{O}(\omega/\mu)  .  
\end{align} 
Going back to (\ref{dappsomewhere}) gives
\begin{align} 
&\text{Im}[\Sigma_{\text{exc}}] = \\& \notag -\frac{\alpha^2 \theta(\omega)}{\pi} \int\displaylimits_{x > \sqrt{\mu}} \hspace{-.8em}d\textbf{x} \ \theta (\omega-(x^2 - \mu)) \hspace{-3em} \int\displaylimits^1_{1-(\omega-(x^2-\mu))/2\beta\mu} \hspace{-3em} dc \ \frac{1}{\sqrt{1-c^2}}  .
\end{align} 
Using that for $0<t<1$: 
\begin{align} 
\label{arccosformula}
\int_{1-t}^1 \frac{1}{\sqrt{1-y^2}} dy = \arccos(1-t) = \sqrt{2t} + \mathcal{O}(t^{3/2}) \ ,
\end{align} 
we obtain 
\begin{align} 
\text{Im}[\Sigma_{\text{exc}}] = -2\alpha^2\theta(\omega)\int_{\sqrt{\mu}}^{\sqrt{\mu + \omega}} x dx \sqrt{\frac{\omega - (x^2 - \mu)}{\beta\mu}}  .
\end{align}
This can be integrated exactly to give: 
\begin{align} 
\label{correctnumapp}
\text{Im}[\Sigma_{\text{exc}}](\omega) = -\frac{2\alpha^2}{3} \frac{1}{\sqrt{\beta\mu}} \cdot\theta(\omega) \omega^{3/2}  .
\end{align} 
The numerical prefactor should be correct, but is of no parametric relevance and is set to unity for convenience, thereby giving formula (\ref{correctnum}) of the main text.

\subsection{FES regime: VB hole recoil} 

In the regime of the FES, not the exciton, but the valence band hole recoils. Near the direct threshold at $\omega = \beta\mu$, the quantity describing the hole decay is Im[$\Sigma_{\rm{VB}}(k_F,\omega)$] as given in (\ref{VBself-energy}), which scales differently compared to the exciton decay because the VB hole has $\Q = k_F$ unlike the $\Q = 0$ exciton (we do not present this computation here since the power law is of not much relevance for the 2DEG absorption we are interested in; see \cite{Pimenov2015} for details). 

\begin{figure}[H]
\center
\includegraphics[width=\columnwidth]{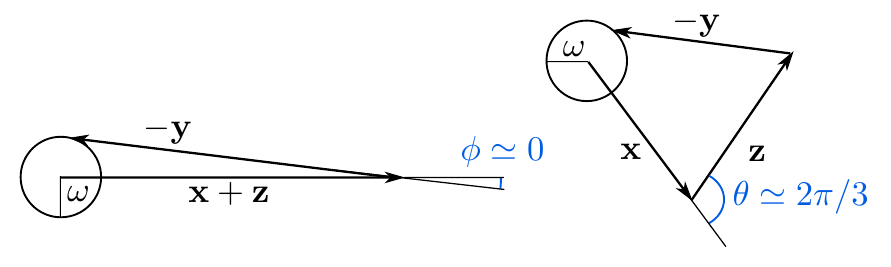}
\caption{Angles contributing to the indirect threshold. The $\omega$-circles indicate smallness in $\omega$, but not the exact power law or prefactor.}
\label{angleom}
\end{figure}

Near the indirect threshold, the VB hole again has momentum $\Q = 0$, and the resulting 2DEG absorption $A(\omega)$ as given in (\ref{Abspowerlaw}) scales as $\sim \omega^3$. This result was already presented in \cite{PhysRevB.35.7551}, though without derivation.
Since the computation is very similar to the previous one for the exciton decay, let us just sketch it: By performing frequency integrals in Figs. \ref{crossed_infmass} and \ref{omega3self}, and momentum substitutions as for the exciton, one arrives at: 
\begin{align}
\label{AIsecond}
&A(\omega) \sim \int_{x^2>\mu}d\textbf{x} \int_{z^2>\mu}d\textbf{z}\int_{y^2<\mu}  d\textbf{y} \\ \nonumber & \delta\!\left(
\omega - \left(x^2 - \mu\right)+\left(y^2 - \mu\right)-\left(z^2 - \mu\right) - \beta\left(\textbf{x} +\textbf{z} - \textbf{y}\right)^2 \right), 
\end{align}
which is similar to the previous expression (\ref{firstsubst}) except for an additional scattering partner, the photoexcited electron (corresponding to the $\textbf{z}$-integral). Again, there can be no cancellations in the deltafunction, and the computation proceeds analogously to sec.\ \ref{excrecsec}. Effectively, the summands $(x^2-\mu), (y^2 - \mu)$ and $(z^2 - \mu)$ contribute a factor of $\omega$ to $A(\omega)$. One factor is fixed by the delta function, such that in total one has $\omega^2$. In addition, there is the hole recoil term $\beta(\textbf{x} + \textbf{z} - \textbf{y})^2$. For this to be of order $\omega$, the angles $\phi = \measuredangle(\textbf{x} + \textbf{z},\textbf{y})$ and $ \theta = \measuredangle(\textbf{x},\textbf{z})$ have to be fixed as depicted in Fig. \ref{angleom}.

The explicit computation shows that each angle restriction give a factor of $\sqrt{\omega}$, such that in total one arrives at $A(\omega)\sim \omega^3$.

\bibliographystyle{prsty_(no_et_al)}

\bibliography{mobile_hole_polaritons_PRB.bib}

\begin{thebibliography}{10}

\bibitem{carusotto2013quantum}
I. Carusotto and C. Ciuti, Rev. Mod. Phys. {\bf 85},  299  (2013).

\bibitem{byrnes2014exciton}
T. Byrnes, N.~Y. Kim, and Y. Yamamoto, Nature Physics {\bf 10},  803  (2014).

\bibitem{PhysRevLett.69.3314}
C. Weisbuch, M. Nishioka, A. Ishikawa, and Y. Arakawa, Phys. Rev. Lett. {\bf
  69},  3314  (1992).

\bibitem{Kasprzak2006}
J. Kasprzak, M. Richard, S. Kundermann, A. Baas, P. Jeambrun, J.~M.~J. Keeling,
  F.~M. Marchetti, M.~H. Szymanska, R. Andre, J.~L. Staehli, V. Savona, P.~B.
  Littlewood, B. Deveaud, and L.~S. Dang, Nature {\bf 443},  409  (2006).

\bibitem{PhysRevLett.104.106402}
F.~P. Laussy, A.~V. Kavokin, and I.~A. Shelykh, Phys. Rev. Lett. {\bf 104},
  106402  (2010).

\bibitem{Anderson1967}
P. Anderson, Phys. Rev. Lett. {\bf 18},  1049  (1967).

\bibitem{PhysRev.163.612}
G.~D. Mahan, Phys. Rev. {\bf 163},  612  (1967).

\bibitem{PhysRev.178.1072}
B. Roulet, J. Gavoret, and P. Nozi\`eres, Phys. Rev. {\bf 178},  1072  (1969).

\bibitem{NOZIERES1969}
P. Nozi\`eres, J. Gavoret, and B. Roulet, Phys. Rev. {\bf 178},  1084  (1969).

\bibitem{PhysRev.178.1097}
P. Nozi\`eres and C. de~Dominicis, Phys. Rev. {\bf 178},  1097  (1969).

\bibitem{combescot1971infrared}
M. Combescot and P. Nozi{\`e}res, J. Phys. {\bf 32},  913  (1971).

\bibitem{PhysRevLett.99.157402}
A. Gabbay, Y. Preezant, E. Cohen, B.~M. Ashkinadze, and L.~N. Pfeiffer, Phys.
  Rev. Lett. {\bf 99},  157402  (2007).

\bibitem{Smolka}
S. Smolka, W.Wuester, F. Haupt, S. Faelt, W. Wegschneider, and A. Imamoglu,
  Science {\bf 346},  332  (2014).

\bibitem{sidler2017fermi}
M. Sidler, P. Back, O. Cotlet, A. Srivastava, T. Fink, M. Kroner, E. Demler,
  and A. Imamoglu, Nature Physics {\bf 13},  255  (2017).

\bibitem{PhysRevB.76.045320}
N.~S. Averkiev and M.~M. Glazov, Phys. Rev. B {\bf 76},  045320  (2007).

\bibitem{PhysRevB.89.245301}
M. Baeten and M. Wouters, Phys. Rev. B {\bf 89},  245301  (2014).

\bibitem{baeten2015mahan}
M. Baeten and M. Wouters, Eur. Phys. J. D {\bf 69},  1  (2015).

\bibitem{gavoret1969optical}
J. Gavoret, P. Nozieres, B. Roulet, and M. Combescot, J. Phys. {\bf 30},  987
  (1969).

\bibitem{PhysRevB.44.3821}
P. Hawrylak, Phys. Rev. B {\bf 44},  3821  (1991).

\bibitem{PhysRevLett.65.1048}
T. Uenoyama and L.~J. Sham, Phys. Rev. Lett. {\bf 65},  1048  (1990).

\bibitem{PhysRevB.35.7551}
A.~E. Ruckenstein and S. Schmitt-Rink, Phys. Rev. B {\bf 35},  7551  (1987).

\bibitem{Nozi`eres1994}
P. Nozi\`eres, J. Phys. I {\bf 4},  1275  (1994).

\bibitem{PhysRevLett.75.1988}
A. Rosch and T. Kopp, Phys. Rev. Lett. {\bf 75},  1988  (1995).

\bibitem{haug1990quantum}
H. Haug and S.~W. Koch, {\em Quantum theory of the optical and electronic
  properties of semiconductors} (World Scientific, Singapore, 2009), Vol.~5.

\bibitem{adhikari1986quantum}
S.~K. Adhikari, Am. J. Phys. {\bf 54},  362  (1986).

\bibitem{PhysRev.93.632}
E. Burstein, Phys. Rev. {\bf 93},  632  (1954).

\bibitem{moss1954interpretation}
T. Moss, Proc. Phys. Soc. London, Sec. B {\bf 67},  775  (1954).

\bibitem{yamamoto1999mesoscopic}
Y. Yamamoto and A. Imamoglu, {\em Mesoscopic quantum optics} (John Wiley \&
  Sons, Inc., New York, 1999), Vol.~1.

\bibitem{PhysRev.153.882}
G.~D. Mahan, Phys. Rev. {\bf 153},  882  (1967).

\bibitem{Betbeder-Matibet2001}
O. Betbeder-Matibet and M. Combescot, Eur. Phys. J. B {\bf 22},  17  (2001).

\bibitem{Note1}
Strictly speaking, this also means $E_B \lesssim \xi $, contradicting Eq.~(\ref
  {scales1}). However, this clearly is a non-universal property, and we will
  not pay any attention to it in the following.

\bibitem{Note2}
In fact, their computation is in 3D, but the case of infinite hole mass is
  effectively 1D anyway.

\bibitem{Combescot2002}
M. Combescot, O. Betbeder-Matibet, and B. Roulet, Europhys. Lett. {\bf 57},
  717  (2002).

\bibitem{combescot2003commutation}
M. Combescot and O. Betbeder-Matibet, Eur. Phys. J. B {\bf 31},  305  (2003).

\bibitem{Combescot2008215}
M. Combescot, O. Betbeder-Matibet, and F. Dubin, Phys. Rep. {\bf 463},  215
  (2008).

\bibitem{5914361120110215}
M. Combescot and O. Betbeder-Matibet, Eur. Phys. J. B {\bf 79},  401   (2011).

\bibitem{suris2001excitons}
R. Suris, V. Kochereshko, G. Astakhov, D. Yakovlev, W. Ossau, J.
  N{\"u}rnberger, W. Faschinger, G. Landwehr, T. Wojtowicz, G. Karczewski, and
  J. Kossut, Phys. Status Solidi B {\bf 227},  343  (2001).

\bibitem{PhysRevB.91.115313}
M. Baeten and M. Wouters, Phys. Rev. B {\bf 91},  115313  (2015).

\bibitem{efimkin2017many}
D.~K. Efimkin and A.~H. MacDonald, Phys. Rev. B {\bf 95},  035417  (2017).

\bibitem{R.Wong1989}
R. Wong, {\em Asymptotic Approximation of Integrals} (Academic Press, Inc., New
  York, 1989).

\bibitem{G.D.Mah2000}
G.D.Mahan, {\em Many-particle-physics}, 3rd ed. (Kluwer Academic/Plenum
  Publishers, New York and London, 2000).

\bibitem{Pimenov2015}
D. Pimenov, Master's thesis, Ludwig Maximilians University Munich, 2015.

\bibitem{Note3}
The regime of $g\protect \qopname \relax o{log}(\beta g^2) \gg 1$ is out of
  reach for the methods used in~\cite {gavoret1969optical}. To study it, a
  consistent treatment of the divergences is needed, similar to~\cite
  {NOZIERES1969}. We will not attempt this here.

\bibitem{Hopfield1969}
J.~J. Hopfield, Comments Solid State Phys. {\bf 2},  40  (1969).

\end{thebibliography}

\end{document}